\documentstyle[12pt]{article}
\title{
States and Observables in Hamiltonian Semiclassical Scalar Electrodynamics
}
\author{{\large  O.Yu.Shvedov} \\
{\it
Sub-Dept. of Quantum Statistics and Field Theory},\\
{\it Dept. of Physics, Moscow State University},\\
{\it 119992, Moscow, Vorobievy Gory, Russia}
}
\date{}
\begin{document}

\def\qp{
\mathrel{\mathop{\bf x}\limits^2},
\mathrel{\mathop{-i\frac{\partial}{\partial {\bf x}}}\limits^1} 
}
\def\gsim{{> \atop \sim}}
\def\lsim{{< \atop \sim}}

\maketitle

\setcounter{page}{0}

\begin{flushright}
hep-th/0412323
\end{flushright}

\section*{Abstract}

The main  notions of semiclassical scalar electrodynamics in different
gauges (Hamiltonian,  Couloumb,  Lorentz)  are  discussed.  These  are
semiclassical states,  Poincare transformations,  fields, observables,
gauge equivalence.  General properties of these objects are formulated
as axioms  of semiclassical theory;  they are heuristically justified.
In particular,  a semiclassical state  may  be  viewed  as  a  set  of
classical background   field   and   quantum  state  in  the  external
background. Superpositions of these "elementary" states  can  be  also
considered. Set  of  all  "elementary"  semiclassical  states  forms a
semiclassical bundle, with base being classical space and fibres being
quantum states   in   the   external   background.   Quantum   symetry
transformations (Poincare  and  gauge  transformations)   are   viewed
semiclassically as automorphisms of the semiclassical bundle. Specific
features of electrodynamics are investigated for different gauges.

%\begin{abstract}
%{\it Keywords:}
%\end{abstract}

%body of a paper here
\footnotetext{e-mail:  shvedov@qs.phys.msu.su}

\footnotetext{This work was supported by the Russian  Foundation  for
Basic Research, project 02-01-01062}

\makeatletter
\@addtoreset{equation}{section}
\makeatother
\renewcommand{\theequation}{\thesection.\arabic{equation}}

\def\lb#1{\label{#1}}
\def\l#1{\lb{#1}}
\def\r#1{(\ref{#1})}
\def\c#1{\cite{#1}}
\def\i#1{\bibitem{#1}}
\def\beq{\begin{equation}}
\def\eeq{\end{equation}}
\def\bez{\begin{displaymath}}
\def\eez{\end{displaymath}}
\def\beb#1\l#2\eeb{\begin{equation} \begin{array}{c} #1 \qquad
\end{array} \label#2  \end{equation}}
\def\bey#1\eey{\begin{displaymath}
\begin{array}{c} #1  \end{array}  \end{displaymath}}

\newpage

\section{Introdcution}

States and  observables  (fields)  are  main  notions of the axiomatic
quantum field theory (QFT).  The difficulty  is  that  it  is  unknown
whether a  nontrivial  model  of  axiomatic AFT exists in 4 dimensions
\c{BLOT}.
All practical calculations in QFT (evaluations of scattering and decay
properties) are  performed  using  the heuristic Lagrangian framework.
All the axioms of QFT are checked within the perturbation theory
\c{BLOT,BS,Z,SF}: if a
formal perturbation series for physical quantities satsfies the set of
axioms then one says that axioms are checked.

Another important heuristic  approximate  method  is  a  semiclassical
approximation.  A  lot  of  examples  of  physical  applicatons of the
semiclassical approximation are known:  these are soliton quantization
theory  \c{soliton,J},  QFT  in a strong external background classical
field  \c{GMM}  or  in  curved   space-time   \c{BD},   the   one-loop
approximation       \c{oneloop},      time-dependent      Hartree-Fock
\c{oneloop,Hartree} and Gaussian approximations \c{Gauss}.

However, the main axiomatic notions (states,  observables and fields),
as well as correspondence princtiple  between  quantum  and  classical
field theories are to be clarified. For the scalar field theories, the
axioms of semiclassical field theory were suggested in \c{Shvedov1}.
The purpose
of this  paper is to formulate and investigate analogs of these axioms
for gauge theories.  Quantum electrodynamics (QED) is considered as an
example of  an  Abelian gauge theory.  Since one knows formulations of
QED in  different  gauges  (Hamiltonian,   Couloumb,   Lorentz),   the
corresponding formulations  of  the  semiclassical  theory  should  be
investigated. One expects all the formulations to be equivalent; these
equivalence should be checked then.

Section 2   deals   with  properties  of  states  and  observables  in
semiclassical field theory.  The discussion is based mostly  on  refs.
\c{Shvedov1,Shvedov2}.
In section  3  different  approaches of quantizing electrodynamics are
reviewed. Section  4  is  devoted to the notion of semiclassical state
for different gauges.  In section  5,  semiclassical  observables  and
transformations are   investigated.   Section  6  contains  concluding
remarks.

\section{Properties of states and observables in semiclassical  field
theory}

Let us  discuss  general  properties  of  semiclassical  field theory.
Consider a simpler example of scalar field theory with the Lagrangian
\beq
{\cal L}   =   \frac{1}{2}   \partial_{\mu}   \varphi   \partial^{\mu}
\varphi - \frac{1}{h} V(\sqrt{h}\varphi),
\l{2.1}
\eeq
$h$ being a small parameter of expansion.

{\bf 1.} A "naive" semiclassical theory can be constructed as follows
(cf. \c{GMM}).
One extracts  a $c$-number component $\Phi(x)/\sqrt{h}$ from the field
$\varphi(x)$:
\beq
\varphi(x) = \frac{\Phi(x)}{\sqrt{h}} + \phi(x);
\l{2.2}
\eeq
then the  remaining part $\phi(x)$ is quantized.  Substitution \r{2.2}
to the Lagrangian \r{2.1} leads to the following action
\bey
I =
\frac{1}{h} \int dx
\left[
\frac{1}{2} \partial_{\mu} \Phi \partial^{\mu} \Phi - V(\Phi)
\right]
+
\frac{1}{\sqrt{h}}
\int dx
\left[
\frac{1}{2} \partial_{\mu} \Phi \partial^{\mu} \phi - V'(\Phi)\phi
\right]
\\ +
\int dx \left[
\frac{1}{2} \partial_{\mu}  \phi  \partial^{\mu}  \phi  -
\frac{1}{2} V'{}'(\Phi)
\phi^2
\right] + ...
\eey
The term of the order $O(1/h)$ is constant and  can  be  omitted;  the
second term  (linear  in $\phi$) vanishes due to classical equation of
motion for $\Phi$; the remaining quadratic term is
\beq
I_2 = \int dx
\left[
\frac{1}{2} \partial_{\mu}  \phi  \partial^{\mu}  \phi  -
\frac{1}{2} V'{}'(\Phi)
\phi^2
\right].
\l{2.3}
\eeq
Then action  \r{2.3}  is  quantized,  and  a  semiclassical  theory is
obtained.

{\bf 2.} The semiclassical theory  can  be  also  constructed  in  the
Hamiltonian approach   as  well.  One  considers  the  quantum  theory
correspoinding to  the  Lagrangian  \r{2.1}   and   investigates   the
semiclassical states  which  depends on the small parameter $h$ due to
the Maslov  substitution  (its  analog  was  suggested   for   quantum
mechanical problems in \c{Maslov1,Maslov2}):
\beq
\Psi \simeq e^{\frac{i}{h} \tilde{S}}
e^{\frac{i}{\sqrt{h}} \int  d{\bf  x} [\Pi({\bf x}) \hat{\varphi}({\bf
x}) -    \Phi({\bf    x})    \hat{\pi}({\bf     x})]}     f     \equiv
\tilde{K}^h_{\tilde{S},\Pi,\Phi} f.
\l{2.4}
\eeq
Here $\hat{\pi}({\bf x})$ is a momentum canonically conjugated to  the
field $\hat{\varphi}({\bf   x})$.   In   the   functional  Schrodinger
representation (the field and momentum operators are
$\hat{\varphi}({\bf x}) = \varphi({\bf x})$, $\hat{\pi}({\bf x}) = - i
\frac{\delta}{\delta \varphi({\bf       x})}$,       states        are
$\Psi[\varphi(\cdot)]$) formula \r{2.4} can be rewritten as
\beq
\Psi[\varphi(\cdot)] = const
e^{\frac{i}{h}S}
e^{\frac{i}{\sqrt{h}} \int  d{\bf  x}  \Pi({\bf  x}) [\varphi({\bf x})
\sqrt{h} - \Phi({\bf x})]}
f[\varphi(\cdot) - \frac{\Phi(\cdot)}{\sqrt{h}}]
\equiv
{K}^h_{S,\Pi,\Phi} f[\varphi(\cdot)]
\l{2.5}
\eeq
with $S=\tilde{S}  +  \frac{1}{2}   \int   d{\bf   x}   \Pi\Phi$.   If
theclassical field  is  $\varphi({\bf  x})  = \Phi({\bf x})/\sqrt{h} +
O(1)$ then the probability amplitude \r{2.5} is large;  otherwise, for
the case   $\varphi   -   \Phi/\sqrt{h}   =   O(1/\sqrt{h})$,   it  is
exponentially small. Therefore, $\Phi({\bf x})/\sqrt{h}$ may be viewed
as a classical component of the field.

The set  of  semiclassical  states  \r{2.5}  may  be treated as {\it a
bundle} ("semiclassical bundle" \c{Shvedov3}),
the base of the  bundle  is  $\{  X
\equiv (S,\Pi(\cdot),\Phi(\cdot))\}$ - a set of classical states;  the
fibres ${\cal F}_X = \{f\}$ are state spaces in given external  fields
$X$. The operator $K_X^h: f\mapsto \Psi$ is called as {\it a canonical
operator.}

{\bf 3.}  A  specific  feature  of   the   Hamiltonian   approach   to
semiclassical field  theory is that ine can investigate also states of
the more  general  form  than  \r{2.5}.  Namely,  one   can   consider
superpositions of the form (cf.\c{MS1})
\beq
\int d\alpha     K^h_{X(\alpha)}     f(\alpha),\qquad     \alpha     =
(\alpha_1,...,\alpha_k),
\l{2.6}
\eeq
which can be viewed as $k$-dimensional surfaces of  the  semiclassical
bundle. Such  superpositions  are  useful  in the soliton quantization
theory due to the well-known problem of zero modes of the solitons. In
quantum mechanics,  one can obtain \c{MS1}
the WKB method and all Maslov methods of \c{Maslov1,Maslov2}
from the wave packet
method with the help of using superpositon \r{2.6}.

One can call  the  state  $K^h_Xf$  as  an  "elementary  semiclassical
state", while  superposition \r{2.6} can be interpreted as a "composed
semiclassical state".

It is necessary to  investigate  the  following  problems  within  the
semiclassical theory:

- action  of  Poincare  transformations ${\cal U}^h_g$ (in particular,
evolution) corresponding to elements $g$ of the Poincare group $G$;

- action of Heisenberg field operators $\hat{\varphi}(x)$;

- inner product of states \r{2.6}.

{\bf 4.} It happens that the following commutation rules are satisfied
as $h\to 0$:
\beb
{\cal U}_g^h K^h_X f\simeq K^h_{u_gX} U_g(u_gX\gets X)f;
\\
\sqrt{h} \hat{\varphi}(x) K^h_X f \simeq
K^h_X [\Phi(x|X) + \sqrt{h}\hat{\phi}(x|X)]f.
\l{2.7}
\eeb
Here $u_g:  X  \mapsto  u_gX$  is a classical Poincare transformation,
$\Phi(x|X)$ is a classical field corresponding to the classical  state
$X$. $\Phi  +  \sqrt{h}  \hat{\phi}$  may be viewed as a semiclassical
field.

An explicit  form  of  the   semiclassical   Poincare   transformation
$U_g(u_gX \gets X)$ was constructed in \c{Shvedov1}.

Important properties  of  classical  and  semiclassical fields for the
model \r{2.1} may be obtained from the Heisenberg equations
\bez
\partial_{\mu}\partial^{\mu} \sqrt{h}\hat{\varphi}(x) +
V'(\sqrt{h}\hat{\varphi}(x)) = 0.
\eez
Making use of \r{2.7}, one finds that
\beb
\partial_{\mu}\partial^{\mu} \Phi(x|X) +
V'(\Phi(x|X)) = 0;\\
\partial_{\mu}\partial^{\mu} \hat{\phi}(x|X) +
V'{}'(\Phi(x|X)) \hat{\phi}(x|X) = 0.
\l{2.7a}
\eeb
Eqs. \r{2.7a} should be completed by the initial conditions at $t=0$:
\bey
\Phi({\bf x}|X)|_{t=0} = \Phi({\bf x}); \quad
\dot{\Phi}({\bf x}|X)|_{t=0} = \Pi({\bf x}); \\
\hat{\phi}({\bf x}|X)|_{t=0} f[\phi(\cdot)] =
\phi({\bf x}) f[\phi(\cdot)];
\qquad
\dot{\hat{\phi}}({\bf x}|X)|_{t=0} f[\phi(\cdot)] =
\frac{1}{i} \frac{\delta}{\delta \phi({\bf x})}
f[\phi(\cdot)].
\eey

Investigate properties  of  Poincare  transformations.  Since operators
${\cal U}_g^h$ should satisfy the group identity
\bez
{\cal U}^h_{g_1g_2} = {\cal U}^h_{g_1} {\cal U}^h_{g_2},
\eez
it follows from relation \r{2.7} that
\beb
u_{g_1g_2} = u_{g_1}u_{g_2}; \\
U_{g_1g_2}(u_{g_1g_2}X \gets X) =
U_{g_1}(u_{g_1g_2}X \gets u_{g_2}X)
U_{g_2}(u_{g_2}X \gets X).
\l{2.8}
\eeb
Properties \r{2.8}  mean  that  the  Poincare  group   acts   on   the
semiclassical bundle as an automorphism group.

For the Poincare transformation $g=(a,\Lambda)$ of the form $x^{\prime
\mu} = \Lambda^{\mu}_{\nu} x^{\nu} + a^{\mu}$,  it  follows  from
Poincare invariance of the fields that
\beq
{\cal U}^h_{g^{-1}}      \hat{\varphi}(x)      {\cal      U}^h_g     =
\hat{\varphi}(w_gx),
\qquad w_gx = \Lambda^{-1}(x-a).
\eeq
therefore, one  obtains  Poincare invariance property of classical and
semiclassical fiedls:
\beb
\Phi(x|u_gX) = \Phi(w_gx|X);\\
\hat{\phi}(x|u_gX) U_g(u_gX \gets X) =
U_g(u_gX \gets X) \hat{\phi}(w_gx|X).
\l{2.10}
\eeb

{\bf 5.} Consider the inner product  $(\Psi,\Psi)$  for  the  composed
semiclassical state $\Psi$ \r{2.6}.  One can calculate it  as  follows
\c{Shvedov2}:
write it as
\beq
(\Psi,\Psi) =            \int             d\alpha             d\alpha'
(K^h_{X(\alpha)}f(\alpha),K^h_{X(\alpha')}f(\alpha')),
\l{2.11}
\eeq
consider the  substitution $\alpha' = \alpha + \sqrt{h}\beta$,  expand
the expression in $\sqrt{h}$.  However,  it is necessary to  write  an
expansion for  the  state $K^h_{X(\alpha + \beta \sqrt{h})} f(\alpha +
\beta \sqrt{h})$ into a series in $\sqrt{h}$.  It can be obtained from
the commutation  rule  between  operators  $ih\frac{\partial}{\partial
\alpha_a}$ and $K^h_{X(\alpha)}$:
\beq
ih \frac{\partial}{\partial \alpha_a}
K^h_{X(\alpha)}f
\simeq
K^h_{X(\alpha)}
\left[
\omega_{X(\alpha)} [\frac{\partial X}{\partial \alpha_a}] -
\sqrt{h}
\Omega_{X(\alpha)} [\frac{\partial X}{\partial \alpha_a}] +
...
\right]
f.
\l{2.12}
\eeq
The $c$-number 1-form $\omega_X[\delta X]$  ('action  form')  and  the
operator-valued 1-form  $\Omega_X[\delta  X]$ (acting in ${\cal F}_X$)
are important objects of the semiclassical theory. Their explicit form
is
\beb
\omega_X[\delta X] = \int d{\bf x} \Pi({\bf x}) \delta \Phi({\bf x}) -
\delta S;\\
\Omega_X[\delta X] f[\phi(\cdot)] =
\int d{\bf x}
[\Pi({\bf x}) \phi({\bf x}) -
\Phi({\bf x})   \frac{1}{i}   \frac{\delta}{\delta   \phi({\bf   x})}]
f[\phi(\cdot)].
\l{2.13}
\eeb
It follows  from the relation
$[ih\frac{\partial}{\partial \alpha_a};
ih\frac{\partial}{\partial \alpha_b}] = 0$ that
the commutator of operators $\Omega$ should be a $c$-number:
\beq
\left[
\Omega_{X(\alpha)} [\frac{\partial X}{\partial \alpha_a}];
\Omega_{X(\alpha)} [\frac{\partial X}{\partial \alpha_b}]
\right] = - i \left\{
\frac{\partial}{\partial \alpha_a}
\omega_{X(\alpha)} [\frac{\partial X}{\partial \alpha_b}]
-
\frac{\partial}{\partial \alpha_b}
\omega_{X(\alpha)} [\frac{\partial X}{\partial \alpha_a}]
\right\}.
\l{2.14}
\eeq
One can  also write relation \r{2.14} in a shorter form.  Namely,  the
$c$-number commutator is related with the  symplectic 2-form $d\omega$:
\beq
[\Omega_X[\delta X_1];\Omega_X[\delta  X_2]]  =  -   id\omega_X(\delta
X_1,\delta X_2).
\l{2.14x}
\eeq
Certainly, commutation relations \r{2.14} and \r{2.14x} are  satisfied
for objects \r{2.13}.

To find an explicit form of $K^h_{X(\alpha + \sqrt{h}\beta)}$, set
\bez
K^h_{X(\alpha + \sqrt{h}\beta)} = K^h_{X(\alpha)} V_h(\alpha,\beta).
\eez
One obtains the following equation on $V_h(\alpha,\beta)$:
\beq
\frac{\partial}{\partial \beta_a}
V_h(\alpha,\beta) \simeq
- \frac{i}{\sqrt{h}}
V_h(\alpha,\beta)
(\omega - \sqrt{h}\Omega)_{X(\alpha + \sqrt{h}\beta)}
[\frac{\partial X}{\partial \alpha_a}(\alpha + \sqrt{h}\beta)].
\l{2.14a}
\eeq
Therefore, in  the  leading order in $h$ one obtains that the operator
$V_h$ is a multiplicator by a rapidly oscillating $c$-number
\bez
V_h(\alpha,\beta) \sim
e^{-\frac{i}{\sqrt{h}} \beta_a    \omega_{X(\alpha)}   [\frac{\partial
X}{\partial \alpha_a}]}.
\eez
The inner product \r{2.11} is taken to the form
\beq
h^{k/2} \int  d\alpha  d\beta  (f(\alpha),V_h(\alpha,\beta)f(\alpha   +
\sqrt{h}\beta)).
\l{2.15}
\eeq
The integrand in \r{2.15} rapidly oscillates,  so that the integral is
exponentially small, except for the special case
\beq
\omega_{X(\alpha)} [\frac{\partial X}{\partial \alpha_a}] = 0.
\l{2.16}
\eeq
Therefore, one should  consider  not  any  superposition  \r{2.6}  but
superpositions obeying {\it the Maslov isotropic condition \r{2.16}.}
Only for  this  case  the  composed   semiclassical   state   is   not
exponentially small.

If condition  \r{2.16}  is  satisfied,  one  can  solve  eq.\r{2.14a},
provided that the commutation relation \r{2.14} is obeyed. One has
\bez
V_h(\alpha,\beta) \sim
e^{{i} \beta_a    \Omega_{X(\alpha)}   [\frac{\partial
X}{\partial \alpha_a}]}.
\eez
Therefore, for $(\Psi,\Psi)$ one has
\beq
(\Psi,\Psi) \simeq
h^{k/2} \int d\alpha d\beta
(f(\alpha),
\prod_a \{
2\pi \delta (\Omega_X[\frac{\partial X}{\partial \alpha_a}])
\} f(\alpha)).
\l{2.16a}
\eeq
One can notice that it is necessary to  multiply  the  composed  state
\r{2.6} by $h^{-k/4}$ in order to satisfy the normalization condition.

Combining commutation  rules  \r{2.7}  and  \r{2.12},  one obtains new
identities:
\beb
\omega_X[\frac{\partial X}{\partial\alpha_a}] =
\omega_{u_gX} [\frac{\partial (u_gX)}{\partial \alpha_a}]; \\
U_g(u_gX\gets X)
\Omega_X[\frac{\partial X}{\partial \alpha_a}] =
\Omega_{u_gX}
[\frac{\partial (u_gX)}{\partial \alpha_a}]
U_g(u_gX\gets X);
\l{2.17}
\eeb
\beq
-i\frac{\partial}{\partial\alpha_a} \Phi(x|X) =
[\hat{\phi}(x|X);\Omega_X[\frac{\partial X}{\partial \alpha_a}]].
\l{2.17a}
\eeq
The first  identity  means  that  the  classical  symplectic structure
should be  invariant  under  Poincare  transformations.   The   second
equality is related with unitarity of Poincare transformations for the
composed states \r{2.16a}.

{\bf 6.} Therefore,  all the problems of semiclassical field theory in
the leading  order  can  be  solved  under  the  following  conditions
("axioms of semiclassical theory"):

{\bf A1.} {\it A semiclassical bundle is given; space of the bundle is
interpreted as a set of semiclassical states; base ${\cal X} = \{X\}$
is a classical space;  elements of fibres  ${\cal  F}_X$  are  quantum
states in a given classical external field $X$.
}

{\bf A2.} {\it
The Poincare   group   $G$  acts  as  an  automorphism  group  on  the
semiclassical bundle; group properties \r{2.8} are satisfied.
}

{\bf A3.}  {\it  Classical  and  semiclassical components of the field
$\Phi(x|X)$ and $\hat{\phi}(x|X)$ are given for all $X \in {\cal X}$.
$\Phi(x|X)$ is  a $c$-number classical field;  $\hat{\phi}(x|X)$ is an
operator distribution acting in ${\cal F}_X$. The property \r{2.10} of
Poincare invariance of the field is satisfied.
}

{\bf A4.} {\it The differential  1-forms  $\omega$  and  $\Omega$  are
given on   $\cal  X$;  $\omega_X[\delta  X]$  is  a  real  $c$-number;
$\Omega_X[\delta X]$ is  an  operator  acting  in  ${\cal  F}_X$.  The
commutation relation \r{2.14} for $\Omega$  and  properties  \r{2.17},
\r{2.17a} for 1-forms, fields and Poincare transformations are satisfied.
}

Therefore, one  can  say that a model of semiclassical field theory is
{\it given in the leading order } if the objects of axioms  A1-A4  are
specified and  their  properties  are  obeyed.  For  the semiclassical
problems, {\it it is not important } whether the "exact" QFT model  is
well-defined mathematically or not.

{\bf 7.} The formulated axioms and properties are not independent.  It
happens that one can express the operator  $\hat{\phi}(x|X)$  via  the
1-form $\Omega$.

Introduce manifestly  covariant  notations.  Let  us identify elements
$X\in {\cal X}$  with  sets  $(S,\overline{\Phi}(x))$
(instead  of  $(S,\Pi({\bf
x}),\Phi({\bf x}))$,  where  $\overline{\Phi}(x)$
is  a  solution  of the Cauchy
problem for classical field equaton
\bez
\partial_{\mu}\partial^{\mu} \overline{\Phi}(x) +
V'(\overline{\Phi}(x)) = 0,
\qquad
\overline{\Phi}|_{t=0} = \Phi({\bf x}),
\quad
\frac{\partial}{\partial t}\overline{\Phi}|_{t=0} = \Pi({\bf x}).
\eez
Then elements  $\delta  X$  of  the tangent space should be identified
with pairs $(\delta S,\delta \overline{\Phi}(x))$,
with $\delta\overline{\Phi}(x)$ being  a
solution of variation equation
\beq
\partial_{\mu}\partial^{\mu} \delta \overline{\Phi}(x)
+ V'{}'(\overline{\Phi}(x)) \delta \overline{\Phi}(x) = 0.
\l{2.18}
\eeq
Then
\bez
\omega[\delta X]  =  \int_{x^0=0}  d{\bf x} \partial_0
\overline{\Phi}({\bf x}) \delta
\overline{\Phi}({\bf x}) - \delta S.
\eez
Property \r{2.14} can be also taken to a manifestly covariant form:
\beb
[\Omega_{\Phi}[\delta_1\overline{\Phi}];\Omega_{\Phi}[\delta_2
\overline{\Phi}]] =
- i \int_{x^0=0} d{\bf x} [\delta_1 \partial_0
\overline{\Phi} \delta_2\overline{\Phi} -
\delta_1 \overline{\Phi} \delta_2 \partial_0 \overline{\Phi}] = \\
- i \int d\sigma^{\mu}
[\partial_{\mu}\delta_1\overline{\Phi} \delta_2\overline{\Phi} -
\delta_1\overline{\Phi} \partial_{\mu} \delta_2\overline{\Phi}].
\l{2.19}
\eeb
Let us    construct    the    operator     $\hat{\phi}(y|X)     \equiv
\hat{\phi}(y|\overline{\Phi})$ from the relation \r{2.17a}:
\beq
-i\delta \overline{\Phi}(y) =
[\hat{\phi}(y|\overline{\Phi}), \Omega_{\Phi}[\delta \overline{\Phi}]].
\l{2.20}
\eeq
One can notice from eq.\r{2.19} that the operator function
\beq
\hat{\phi}(y|\overline{\Phi}) = \Omega_{\Phi}[\delta \overline{\Phi}^{(y)}]
\l{2.21}
\eeq
satisifes eq.\r{2.20}, provided that $\delta \overline{\Phi}^{(y)}$
is a solution
of eq.\r{2.18}, such that the additional conditions of the form
\bez
\delta \overline{\Phi}^{(y)}_{x^0=y^0} = 0,
\qquad
\delta \partial_0 \overline{\Phi}^{(y)}_{x^0=y^0} = \delta({\bf x}-{\bf y}).
\eez
are satisfied.
One can  also  express  $\delta  \overline{\Phi}^{(y)}(x)$
via the retarded Green
function for eq.\r{2.18}:
\bey
[\partial_{\mu}\partial^{\mu} + V'{}'(\overline{\Phi}(x))]
D^{ret}_{\Phi}(x,y)  =
\delta(x,y);\\
D^{ret}_{\Phi}(x,y) = 0, \quad x < y,
\eey
since
\bez
\delta \overline{\Phi}^{(y)}(x) =D^{ret}_{\Phi}(x,y), \quad x > y.
\eez

If the   definition  \r{2.21}  is  accepted,  properties  of  Poincare
invariance of fields are corollaries of  properties  of  the  operator
$\Omega$. The  1-forms  seems then to be more important objects of the
semiclassical theory than fields.

{\bf 8.} Let us discuss now general specific features of semiclassical
gauge field theories.  It happens that some of classical states may be
{\it gauge-equivalent}:  $X_1 \sim X_2$ \c{Shvedov4}.
This means that semiclassical
states $K^h_{X_1}f_1$  and  $K^h_{X_2}f_2$  approximately  coincide as
$h\to 0$
\beq
K^h_{X_1}f_1 \simeq K^h_{X_2}f_2
\l{2.22}
\eeq
under condition
\bez
f_2 = V(X_2\gets X_1)f_1, \qquad X_2\sim X_1
\eez
for some unitary operator $V(X_2\gets X_1)$.
Let us investigate its properies. It is obvious that
\beb
X_1\sim X_2, X_2\sim X_3 \Rightarrow X_1 \sim X_3;\\
V(X_3\gets X_1) = V(X_3\gets X_2) V(X_2\gets X_1).
\l{2.23}
\eeb
Further, it follows from eq.\r{2.22} that
${\cal U}_g^h K^h_{X_1} f_1 \simeq {\cal U}_g^h K^h_{X_2} f_2$, so that
\beb
X_1 \sim X_2 \Rightarrow u_gX_1 \sim u_gX_2;\\
V(u_gX_2\gets u_gX_1) U_g(u_gX_1 \gets X_1) =
U_g(u_gX_2 \gets X_2) V(X_2\gets X_1).
\l{2.24}
\eeb
If quantum  field  operators   (such   as   vector   potential)   were
well-defined for gauge theories, the relation
$\sqrt{h}\hat{\varphi}(x) K^h_{X_1}f_1 \simeq
\sqrt{h}\hat{\varphi}(x) K^h_{X_2}f_2$ would imply that
\beb
\Phi(x|X_2) = \Phi(x|X_1); \\
\hat{\phi}(x|X_2) V(X_2\gets X_1) = V(X_2\gets X_1) \hat{\phi}(x|X_1),
\quad X_1 \sim X_2.
\l{2.24a}
\eeb
Finally, let $(X_i,f_i)$ be $\alpha$-dependent.  Let us  differentiate
relation \r{2.22};
$ih \frac{\partial}{\partial\alpha_a} K^h_{X_1}f_1
\simeq
ih \frac{\partial}{\partial\alpha_a} K^h_{X_2}f_2$. One obtains
\beb
\omega_{X_1}[\frac{\partial X_1}{\partial\alpha_a}] =
\omega_{X_2}[\frac{\partial X_2}{\partial\alpha_a}]; \\
V(X_2\gets X_1) \Omega_{X_1}[\frac{\partial X_1}{\partial \alpha_a}] =
\Omega_{X_2}[\frac{\partial X_2}{\partial\alpha_a}]
V(X_2\gets X_1),
\quad X_1(\alpha) \sim X_2(\alpha).
\l{2.25}
\eeb
In particular, relations \r{2.25} imply that
\bez
\omega_X[\delta X] = 0,
\quad
\Omega_X[\delta X] = 0
\mbox{ if } X+\delta X \sim X.
\eez
Therefore, for  gauge  theories  an  additional axiom of semiclassical
theory (concerning $V(X_2\gets X_1)$) should be formulated.

{\bf 9.} It happens that axioms A2 and A3 should be  revised.  Namely,
classical states $u_{g_1g_2}X$ and $u_{g_1}u_{g_2}X$ may be equivalent
but not equal. This means that property \r{2.8} should be rewritten as
\beb
u_{g_1g_2} X \sim u_{g_1} u_{g_2}X, \\
U_{g_1g_2} (u_{g_1g_2}X \gets X) =
V(u_{g_1g_2}X \gets u_{g_1}u_{g_2}X)
U_{g_1} (u_{g_1}u_{g_2}X \gets u_{g_2}X)
U_{g_2} (u_{g_2}X \gets X).\\
\l{2.26}
\eeb
Axiom A3  also  reqiure  a  revision since the vector potential ${\cal
A}^{\mu}(x)$ is {\it not } an observable.  It is  more  convenient  to
consider {\it gauge-invariant} observables
\bez
\hat{O} = O[\sqrt{h} \hat{\varphi}(\cdot)].
\eez
An analog of commutation rule \r{2.7} will be written as
\beq
\hat{O} K^h_X f \simeq K^h_X [O(X) + \sqrt{h} \Xi O(X) + ...] f.
\l{2.27}
\eeq
We see that for gauge theories one should assign a  c-number  quantity
$O(X)$ and  an  operator $\Xi O(X)$ to each gauge-invariant functional
$O[\Phi(\cdot)]$.

Note that for the scalar case
\beq
O(X) = O[\Phi(\cdot|X)];
\quad
\Xi O(X) = \int dx \frac{\delta O}{\delta \Phi(x)} \hat{\phi}(x|X).
\l{2.28}
\eeq

Investigate general properties of the infinitesimal objects.  First of
all, write the Poincare invariance property
\beq
{\cal U}^h_{g^{-1}} O[\sqrt{h} \hat{\varphi}(\cdot)] {\cal U}^h_g
= O[\sqrt{h}           v_g           \hat{\varphi}(\cdot)]           =
(v_gO)[\sqrt{h}\hat{\varphi}(\cdot)].
\l{2.29}
\eeq
Here for scalar and vector fields
\bez
v_g\hat{\varphi}(\cdot) \equiv \hat{\varphi}(w_g\cdot);
\quad
v_g\hat{A}^{\mu}(\cdot) \equiv \Lambda^{\mu}_{\nu}
\hat{A}^{\nu}(w_g\cdot);
\quad
w_g x = \Lambda^{-1}(x-a).
\eez
Making use of relation \r{2.27}, one finds that
\beb
O(u_gX) = (v_gO)(X); \\
\Xi O(u_gX) U_g(u_gX \gets X) = U_g(u_gX \gets X) \Xi(v_gO)(X).
\l{2.30}
\eeb

Next, obtain an analog of relation \r{2.17a}. Let $X=X(\alpha)$. Apply
the differential  operator  $ih\frac{\partial}{\partial  \alpha_a}$ to
relation \r{2.27}. making use of eq.\r{2.12}, one obtains
\bez
[(O+\sqrt{h}\Xi O + ...)(X);
(\omega -   \sqrt{h}\Omega   +   ...)_X   [\frac{\partial  X}{\partial
\alpha_a}]] = ih \frac{\partial}{\partial \alpha_a}
(O+\sqrt{h}\Xi O + ...)(X).
\eez
In the leading order in $\sqrt{h}$,
\beq
[(\Xi O)(X), \Omega_X [\frac{\partial X}{\partial \alpha_a}]] =
- i \frac{\partial O(X)}{\partial \alpha_a}.
\l{2.31}
\eeq
Eq.\r{2.31} can be also rewritten in terms of differential forms:
\beq
[(\Xi O)(X), \Omega_X[\delta X]] = - idO(\delta X).
\l{2.32}
\eeq
The operator $(\Xi O)(X)$ can be expressed via $\Omega_X$.  Namely, if
it is looked for in the form
\beq
(\Xi O)(X) = - \Omega_X [\nabla_OX],
\l{2.33}
\eeq
the comutation relation \r{2.32} will take the form
\beq
d\omega(\cdot,\nabla_OX) = dO.
\l{2.34}
\eeq
One should  investigate the problem of solvability of eq.\r{2.34}.  It
happens that the solution $\nabla_OX$ of \r{2.34} is  found  up  to  a
vector $\delta  X_0$ such that $\omega_X[\delta X_0] = 0$.  Therefore,
the operator $\Omega_X[\nabla_OX]$ is well-defined.

To justify relation \r{2.33} up to a c-number,
one should check that any operator commuting with all $\Omega_X[\delta
X]$ is a c-number. This is a correct statement for electrodynamiics.

Finally, obtain an analog of eq.\r{2.24a}.  Since $\hat{O}K^h_{X_1}f_1
\simeq \hat{O}K^h_{X_2} f_2$ under conditions \r{2.22}, one has
\beq
O(X_2) = O(X_1);\\
\Xi O(X_2) V(X_2\gets X_1) = V(X_2\gets X_1) \Xi O(X_1),
\quad X_1 \sim X_2.
\l{2.35}
\eeq
Thus, for gauge theories one should reformulate axioms A2,  A3, A4 and
formulate a new axiom A5.

{\bf A2'}.   {\it   For  each  Poincare  transformation  $g\in  G$,  a
transformation $u_g:  {\cal X} \to {\cal X}$ and an  unitary  operator
$U_g(u_gX\gets X):  {\cal  F}_X  \to  {\cal F}_{u_gX}$ are given.  The
property \r{2.26} is satisfied.
}

{\bf A3'}.  {\it  Let  $O[\Phi(\cdot)]$ be a gauge-invariant classical
functional of  fields  $\Phi$.  Then   classical   and   semiclassical
components ($O(X)$ and $\Xi O(X)$) of the quantum observable are given
for all $X\in \cal X$. $O(X)$ is a c-number classical observable, $\Xi
O(X)$ is  an  operator  in  ${\cal  F}_X$.  The  property  \r{2.30} of
Poincare invariance of observables is satidfied.
}

{\bf A4'}: eq.\r{2.17a} should be substituted be eq.\r{2.31}.

{\bf A5.} {\it
An equivalence  relation  on the base $\cal X$ is given.  For any pair
classically equivalent  states  $X_1\sim  X_2$  an  unitary   operator
$V(X_2\gets X_1):  {\cal F}_{X_1} \to {\cal F}_{X_2}$ is specified. It
satisfies properties \r{2.23}, \r{2.24}, \r{2.25} and \r{2.35}.
}

Let us  check  now  the expectations of this section for semiclassical
gauge theories.  Scalar electrodynamics is a simple example  of  gauge
theory. First,  review  the  main approaches to quantize the theory in
Hamiltonian, Couloumb  and  Lorentz  gauges.  Then  the  semiclassical
approximation will be developed.

\section{Quantization of scalar electrodynamics (Hamoltonian, Couloumb
and Lorentz gauges)}

There are different ways to quantize gauge theories.  One can use  the
Dirac approach \c{D}
or the manifestly covariant BRST-BFV quantization
\c{BRST,BFV}.  Let
us review these approaches for the scalar electrodynamics - a model 
specifying interaction of the complex scalar field $\theta$ with
electromagnetic field $A^{\mu}$. To simplify notations, set
$(A^{\mu},\theta,\theta^*) \equiv \varphi$.

\subsection{Dirac quantization}

{\bf 1.} One starts from the Lagrangian of the form
\beq
{\cal L} = D_{\mu} \theta^* D^{\mu} \theta -  m^2\theta^*\theta  -
\frac{1}{4} F_{\mu\nu}        F^{\mu\nu}        -        \frac{1}{h}
V(h\theta^*\theta).
\l{3.1}
\eeq
Here
$F_{\mu\nu}   =
\partial_{\mu} A_{\nu}   -  \partial_{\nu}  A_{\mu}$,
$D_{\mu}   =   \partial_{\mu}   -   i\sqrt{h}A_{\mu}$
is  a  covariant
derivative (electric charge is set to be $\sqrt{h}$ for simplification
of notations),  $A_{\mu}$  is  a
vector potential,  $\theta$  is a scalar field of mass $m$,  $V$ is a
self-interaction potential  of  the  scalar  field. The
momenta canonically conjugated to $A^{\mu}$, $\theta$ and $\theta^*$
are
\bez
E_{\mu} = \frac{\partial {\cal L}}{\partial \dot{A}^{\mu}}= F_{\mu 0},
\qquad
\pi_{\theta}=\frac{\partial {\cal  L}}{\partial  \dot{\theta}^*} = D_0\theta,
\qquad
\pi^*_{\theta} =    \frac{\partial    {\cal   L}}{\partial   \dot{\theta}}   =
D_0\theta^*,
\eez
so that $E_0=0$. The Hamiltonian has the form
\bez
H = \int d{\bf x} [{\cal H}({\bf x}) + A_0({\bf x})
{\sf \Lambda}_{\bf x}]
\eez
with the Hamiltonian density
\beq
{\cal H}({\bf x})
= \frac{1}{2} E_kE_k + \frac{1}{4} F_{ij} F_{ij} +
\pi_{\theta}^*\pi_{\theta} +
D_i\theta^* D_i\theta    +   m^2\theta^*\theta   +   \frac{1}{h}
V(h\theta^*\theta)
\l{3.2}
\eeq
and constraints
\beq
{\sf \Lambda}_{\bf x} = \partial_kE_k + i \sqrt{h} (\pi_{\theta}^*\theta -
\pi_{\theta}\theta^*).
\l{3.3}
\eeq
The $A_0$-component of the vector potential appears to be
a Lagrange multiplier.

{\bf 2.}
Consider the   quantum   theory   in   the    functional    Schrodinger
representation. States of the system are specified as functionals
$\Psi[A^k(\cdot),\theta(\cdot),\theta^*(\cdot)]
\equiv \Psi[\varphi(\cdot)]$. The field
operators
$\hat{\varphi}({\bf x}) \equiv
(\hat{A}^k({\bf x}),\hat{\theta}({\bf x}),\hat{\theta}^*(\cdot))$
are multiplicators  by  $A^k({\bf   x})$,   $\theta({\bf   x})$   and
$\theta^*({\bf x})$, while the momenta operators
$\hat{\pi}({\bf x}) \equiv
(\hat{E}_k({\bf x}),\hat{\pi}_{\theta}({\bf
x}),\hat{\pi}_{\theta}^*({\bf x}))$
are
\beq
\hat{E}_k = -i \frac{\delta}{\delta A^k({\bf x})}, \qquad
\hat{\pi}_{\theta}({\bf x}) = -i \frac{\delta}{\delta \theta^*({\bf x})},
\qquad
\hat{\pi}_{\theta}^*({\bf x}) = -i \frac{\delta}{\delta \theta({\bf x})}.
\l{3.3-}
\eeq
The quantum operator
$\hat{\cal  H}({\bf  x})$  corresponding  to the
classical Hamiltonian  density ${\cal H}({\bf x})$ \r{3.2} is obtained
from expression \r{3.2} by substituting classical variables  by  their
quantum analogs \r{3.3-}, while
\beq
\hat{P}^0 = \int d{\bf x} \hat{\cal  H}({\bf  x})
\l{3+0}
\eeq
is quantum Hamiltonian.
The Schrodinger equation for the time-dependent states
$\Psi^t[A^k(\cdot),\theta(\cdot),\theta^*(\cdot)]$ reads
\beq
i \dot{\Psi}^t = \hat{P}^0 \Psi^t.
\l{3.3a}
\eeq

{\bf 3.}
There are several ways to take the constraints into account.

In the original Dirac approach \c{D},
physical states $\Psi^t_D$ satisfy not only equation \r{3.3a} but also
the additional conditions
\beq
\hat{\Lambda}_{\bf x} \Psi_D^t = 0
\l{3.4}
\eeq
The operators $\hat{\Lambda}_{\bf x}$ are
quantum analogs of constraints \r{3.3},
\beq
\hat{\Lambda}_{\bf x}  =  \partial_k  \frac{1}{i} \frac{\delta}{\delta
A^k({\bf x})} + \sqrt{h} \left(
\theta({\bf x}) \frac{\delta}{\delta \theta({\bf x})}
- \theta^*({\bf x}) \frac{\delta}{\delta \theta^*({\bf x})}
\right).
\l{3.4a}
\eeq
Since
\beq
[ \hat{\Lambda}_{\bf x}; \hat{P}_0 ] = 0,
\l{3.4b}
\eeq
condition \r{3.4}  conserves under time evolution.  The most difficult
problem in the original  Dirac  approach  is  to  introduce  an  inner
product.

One   can  perform  the  Couloumb  gauge  quantization.  The  wave
functionals $\Psi_C$ are considered on the surface
\beq
\partial_k A^k({\bf x}) = 0
\l{3-1}
\eeq
only. They depend on $A^k_{\perp}$, $\theta$, $\theta^*$ then,
\bez
\Psi_C = \Psi_C [A_{\perp}, \theta,\theta^*],
\eez
where
\beq
A_{\perp}^k({\bf x}) = (\delta_{kl} -
\frac{\partial_k\partial_l}{\partial^2})A^l({\bf x});
\l{3.i1}
\eeq
so that
\beq
A^k({\bf x})       =       (\delta_{kl}       -       \frac{\partial_k
\partial_l}{\partial^2}) A_{\perp}^l({\bf  x}) + \partial_k\gamma({\bf
x}),
\qquad
\gamma({\bf x}) = \frac{1}{\partial^2} \partial_l A^l({\bf x}).
\l{3.i0}
\eeq
The operators  $\hat{A}^k({\bf x})$ and $\hat{E}_k({\bf x})$ should be
rewritten in the Couloumb gauge in the following way.  Since $\Psi$ is
viewed on  the  surface  \r{3-1},  one  has  $\hat{A}^k  ({\bf  x})  =
A^k_{\perp}({\bf x})$. One also has
$$
\frac{1}{i} \frac{\delta}{\delta A^k({\bf x})} =
\left(
\delta_{kl} - \frac{\partial_k \partial_l}{\partial^2}
\right)
\frac{1}{i} \frac{\delta}{\delta A^l_{\perp} ({\bf x})}
-       \frac{1}{\partial^2}       \partial_l      \frac{1}{i}
\frac{\delta}{\delta \gamma({\bf x})},
$$
provided that the continuation of $\Psi_C$ for arbitrary $A^k$ is given.
If condition \r{3.4} is satisfied, one has
\bez
\left[
\frac{\delta}{\delta \gamma({\bf x})} + i\sqrt{h}
\left(
\theta({\bf x}) \frac{\delta}{\delta \theta({\bf x})}
- \theta^*({\bf x}) \frac{\delta}{\delta \theta^*({\bf x})}
\right)
\right] \Psi_C = 0.
\eez
Therefore, in the Couloumb gauge the field operators are
\beb
\hat{E}_k^{(C)} =
\left( \delta_{kl} - \frac{\partial_k \partial_l}{\partial^2}
\right) \frac{1}{i} \frac{\delta}{\delta A^l_{\perp} ({\bf x})}
- \sqrt{h} \frac{1}{\partial^2} \partial_k
\left(
\theta({\bf x}) \frac{\delta}{\delta \theta({\bf x})}
- \theta^*({\bf x}) \frac{\delta}{\delta \theta^*({\bf x})}
\right);
\\
\hat{A}^k({\bf x}) = A^k_{\perp}({\bf x}).
\l{3-2}
\eeb
The quantum  Hamiltonian  density  $\hat{\cal H}({\bf x})$ is obtained
from expression \r{3.2} by substituting classical variables by quantum
analogs \r{3-2}, while quantum Hamiltonian is of the form \r{3+0}. The
inner product in the Couloumb gauge is
\beq
<\Psi_C|\Psi_C> =      \int      DA_{\perp}      D\theta^*       D\theta
|\Psi_C[A_{\perp},\theta,\theta^*]|^2.
\l{3-3}
\eeq

{\bf 4.} An  alternative  way  to  quantize  gauge  theories  is to use the
refined algebraic quantization  approach \c{Marolf}.
States will be denoted as $\Psi_H$ ("Hamiltonian gauge").
It  is  the  most  suitable
quantization for semiclassical approximation.  Instead of imposing the
constraints on physical states,  one modifies the inner product of the
theory \c{inner},
\beb
<\Psi_H,\Psi_H> =  (\Psi_H,  \prod_{\bf  x}
\delta(\hat{\Lambda}_{\bf x}) \Psi_H) =
\\
\int DA D\theta D\theta^* (\Psi_H[A,\theta,\theta^*])^*
\prod_{\bf  x}  \delta(\Lambda_{\bf x}) \Psi_H[A,\theta,\theta^*].
\l{3.5}
\eeb
Because of  eq.\r{3.4b},  the inner product \r{3.5} is invariant under
time evolution.

Note that the inner product \r{3.5} is degenerate. For example, states
of the form
\beq
\int d{\bf x} \alpha({\bf x}) \hat{\Lambda}_{\bf x} Y; \qquad
(\exp (\frac{i}{\sqrt{h}} \int d{\bf x}  \alpha({\bf  x})  \hat{\Lambda}_{\bf
x}) -1 ) Y
\l{3.5a}
\eeq
are of zero norm.  Thus,  we should say that state functionals  $\Psi_H$
and $\Psi_H'$ are equivalent if their difference is of zero norm,
\bez
\Psi_H \sim \Psi_H' \equiv <\Psi_H - \Psi_H', \Psi_H- \Psi_H'> = 0.
\eez
The corresponding factorspace is viewed as a physical state space.
Thus, quantum states
\beq
\Psi_H \sim
\exp (\frac{i}{\sqrt{h}} \int d{\bf x}  \alpha({\bf  x})  \hat{\Lambda}_{\bf
x}) \Psi_H
\l{3.5b}
\eeq
may be viewed as gauge equivalent.

Relationship between states $\Psi_H$ and $\Psi$ in the Dirac and refined
algebraic quantization approaches is as follows (cf. \c{Shvedov5}),
\beq
\Psi_D = \prod_{\bf x} \delta (\hat{\Lambda}_{\bf x}) \Psi_H.
\l{3.6}
\eeq
We notice that condition \r{3.4}  is  automatically  satisfied,  while
equivalent $\Psi_H$-states give the same $\Psi$-state.

An explicit    form    of   the   operator   $\prod_{\bf   x}   \delta
(\hat{\Lambda_{\bf x}})$ can be written via the  following  functional
integral,
\beq
\prod_{\bf x}   \delta(\hat{\Lambda}_{\bf   x})   =   \int   D\alpha  \exp[-
\frac{i}{\sqrt{h}}
\int d{\bf x} \alpha({\bf x}) \hat{\Lambda}_{\bf x}].
\l{3.6c}
\eeq
One also has
\bey
\exp[-\frac{i\tau}{\sqrt{h}}
\int  d{\bf  x} \alpha({\bf x}) \Lambda_{\bf x}]
\Psi_H[A^k(\cdot),\theta(\cdot),\theta^*(\cdot)] =
\\
\Psi_H[A^k(\cdot) +     \frac{\tau}{\sqrt{h}}     \partial_k
\alpha(\cdot),
\theta(\cdot) e^{-i\tau        \alpha(\cdot)},       \theta^*(\cdot)
e^{i\tau\alpha(\cdot)}],
\eey
since both left-hand and right-hand sides of this  relation  obey  the
same equation
\bez
i\sqrt{h}
\frac{\partial   \Psi_H^{\tau}}{\partial   \tau}   =  \int  d{\bf  x}
\alpha({\bf x}) \hat{\Lambda}_{\bf x} \Psi_H^{\tau}
\eez
and initial condition.
We see  that operator \r{3.6c} generates a gauge transformation on the
configuration field space.
Therefore,  formulas \r{3.5},  \r{3.6} can  be
written as
\beq
(\Psi_H,\Psi_H) =
\int DA   D\theta^*   D\theta   D\alpha  \Psi_H^*[A,\theta,\theta^*]
\Psi_H[A+\frac{1}{\sqrt{h}}\partial\alpha,
\theta    e^{-i\alpha},     \theta^*
e^{i\alpha}];
\l{3.9}
\eeq
\beq
\Psi_D[A,\theta,\theta^*] = \int D\alpha \Psi_H [A + \frac{1}{\sqrt{h}}
\partial\alpha, \theta e^{-i\alpha}, \theta^* e^{i\alpha}].
\l{3.i4}
\eeq
Let us  perform  the  linear change of variables \r{3.i0} and consider
the functional  $\Psi_D[A_{\perp},\gamma,\theta,\theta^*]$.  Then  the
obtained formulas will be taken to the form
\bey
\Psi_D[A_{\perp},\gamma,\theta,\theta^*] =         \int        D\alpha
\Psi_H[A_{\perp}, \gamma+  \frac{1}{\sqrt{h}}  \alpha,  \theta   e^{-i\alpha},
\theta^* e^{i\alpha}];\\
<\Psi_H|\Psi_H> =
\int DA_{\perp}      D\gamma      D\theta^*      D\theta     D\alpha
\Psi_H^*[A_{\perp},\gamma,\theta^*,\theta]
\Psi_H[A_{\perp},\gamma+\frac{1}{\sqrt{h}}\alpha,
\theta e^{-i\alpha},
\theta^*e^{i\alpha}] = \\
\int      DA_{\perp}      D\theta^*       D\theta
|\Psi[A_{\perp},0,\theta,\theta^*]|^2.
\eey
Formula \r{3-3} is then justified.

{\bf 5.} Any    Poincare
transformation $(a,\Lambda)$,
\bez
x^{\mu \prime} = \Lambda^{\mu}_{\nu} x^{\nu} + a^{\mu}
\eez
is a composition of time and space translations,  boost
and spatial rotations,
\beq
(a,\Lambda) = (a^0,1) ({\bf a},1) (0, \exp(\alpha^k l^{0k}))
(0, \exp(\frac{1}{2} \theta_{sm} l^{sm}))
\l{3|1}
\eeq
with $\theta_{sm} = - \theta_{ms}$,
\bez
(l^{\lambda\mu})^{\alpha}_{\beta} =        -         g^{\lambda\alpha}
\delta^{\mu}_{\beta} + g^{\mu\alpha} \delta^{\lambda}_{\beta}.
\eez
The operator   $\hat{\cal  U}^h_{a,\Lambda}$  of  the  quantum  Poincare
transformation is
\beq
\hat{\cal U}^h_{a,\Lambda} = \exp[i\hat{P}^0a^0] \exp[-i\hat{P}^ja^j]
\exp[i\alpha^k \hat{M}^{0k}]    \exp[\frac{i}{2}   \hat{M}^{lm}
\theta_{lm}].
\l{3|2}
\eeq

The operator $\hat{P}^0$ has been already defined  (formula  \r{3+0}),
while
\beb
\hat{P}^l = \int d{\bf x} \hat{\cal P}^l({\bf x}); \quad
\hat{M}^{k0} = \int d{\bf x} x^k \hat{\cal H} ({\bf x}); \\
\hat{M}^{kl} =  \int  d{\bf  x}  [x^k  \hat{\cal  P}^l({\bf  x}) - x^l
\hat{\cal P}^k({\bf x}) + \hat{E}_l \hat{A}^k - \hat{E}_k \hat{A}^l].
\l{3|3}
\eeb

Here the operators
$\hat{\cal H}({\bf x})$,  $\hat{\cal P}^l({\bf x})$, $\hat{E}_l$,
$\hat{A}^l$ are obtained from eq.\r{3.2} and expression
\bez
{\cal P}^l = - \partial_l \theta \pi^* - \partial_l \theta^*  \pi  -
\partial_l A^s E_s
\eez
by substituting   classical   variables   by  their  quantum  analogs,
eqs.\r{3.3-} and  \r{3-2}  for   Hamiltonian   and   Couloumb   gauges
correspondingly.

Since Poincare  generators  commute with constraints on the constraint
surface, condition  \r{3.4},  inner  product  \r{3.5}  conserve  under
Poincare transformations. The Poincare algebra infinitesimal relations
\beb
[ \hat{P}^{\lambda},\hat{P}^{\mu} ] = 0,
\quad
\left[ \hat{M}^{\lambda\mu},\hat{P}^{\sigma} \right] =
i(g^{\mu\sigma}\hat{P}^{\lambda} - g^{\lambda\sigma}\hat{P}^{\mu});
\\
\left[ \hat{M}^{\lambda\mu},\hat{M}^{\rho\sigma} \right] =
-i(g^{\lambda\rho}\hat{M}^{\mu\sigma}
- g^{\lambda\sigma} \hat{M}^{\mu\rho}
+ g^{\mu \sigma} \hat{M}^{\lambda \rho}
- g^{\mu \rho} \hat{M}^{\lambda \sigma})
\l{3|4a}
\eeb
are satisfied  on  the  constraint  surface.  This  implies  that  the
operators $\hat{\cal U}^h_{a,\Lambda}$ indeed form a  representation  of
the Poincare group.

\subsection{The Gupta-Bleuler (BRST-BFV) approach}

The manifestly covariant quantization technique \c{BRST,BFV}
(quantization in Lorentz gauge) of electrodynamics is as follows
(see, for example, \c{Fulop}).  States
are specified as functionals
\beq
\Psi_L [A^0(\cdot),A^k(\cdot),\theta(\cdot),\theta^*(\cdot)].
\l{3.i4a}
\eeq
An indefinite inner product is introduced:
\beb
<\Psi_L,\Psi_L> =  \\
\int   DA^k   D\lambda   D\theta   D\theta^*
(\Psi_L[A^k(\cdot),-i\lambda(\cdot),\theta(\cdot),\theta^*(\cdot)])^*
\Psi_L[A^k(\cdot),i\lambda(\cdot),\theta(\cdot),\theta^*(\cdot)]
\l{3.i4m}
\eeb
The Gupta-Bleuler constraint condition is imposed on physical states:
\beq
[\frac{1}{i} \frac{\delta}{\delta        A^0({\bf        x})}        -
\frac{i}{\sqrt{-\Delta}} \Lambda_{\bf x}] \Psi_L = 0.
\l{3.i5}
\eeq
Moreover, states
\beq
\Psi_L \sim \Psi_L + \int d{\bf x} \beta({\bf x})
[\frac{1}{i} \frac{\delta}{\delta        A^0({\bf        x})}        +
\frac{i}{\sqrt{-\Delta}} \Lambda_{\bf x}] Y_L
\l{3.i5a}
\eeq
are set to be equivalent. Definition of equivalence realtion \r{3.i5a}
is reasonable since state
$\int d{\bf   x}   \beta({\bf  x})  [\frac{1}{i}  \frac{\delta}{\delta
A^0({\bf x})} + \frac{i}{\sqrt{-\Delta}} \Lambda_{\bf x}] Y_L$
is orthogonal to any physical state.

One can notice that this approach is equivalent to the Dirac approach.
Condition \r{3.i5} implies that
\beq
\Psi_L[A^k(\cdot),A^0(\cdot),\theta(\cdot),\theta^*(\cdot)]      =
\exp[-\int d{\bf x} A^0({\bf x}) \frac{1}{\sqrt{-\Delta}} \Lambda_{\bf
x}] \Psi_H[A^k(\cdot),\theta(\cdot),\theta^*(\cdot)].
\l{3.i6}
\eeq
The inner product \r{3.i4m} will be rewritten then as
\bez
\int DA^k                      D\theta                     D\theta^*
\Psi_H^*[A^k(\cdot),\theta(\cdot),\theta^*(\cdot)] \int       D\lambda
e^{-2i\int d{\bf     x}     \lambda_{\bf     x}
\frac{1}{\sqrt{-\Delta}}
\Lambda_{\bf    x}}
\Psi_H[A^k(\cdot),\theta(\cdot),\theta^*(\cdot)]
\eez
We come to formula  \r{3.5}  up  to  a  field-independent  normalizing
factor.

The Poincare generators are given by formulas \r{3+0}, \r{3|3} with
\bey
\hat{\cal H}_L  =  \hat{\cal  H}  +  \hat{A}_0 \hat{\Lambda}_{\bf x} -
\frac{\xi}{2} \hat{E}_0^2 + \hat{A}^k \partial_k \hat{E}_0, \\
\hat{\cal P}_L^l = \hat{\cal P}^l - \partial_l \hat{A}^0 \hat{E}_0.
\eey
Here $\hat{\cal  H}$  and $\hat{\cal P}^l$ are Hamiltonian and momenta
densities for the Hamiltonian gauge, $\hat{E}_0({\bf x}) = \frac{1}{i}
\frac{\delta}{\delta A^0({\bf  x})}$,  $\xi$ is a real parameter.  The
algebraic properties  \r{3|4a}
are satisfied exactly (not only on the constraint surface).
Making use of eq.\r{3.i6},  we find that equations  of
motion for Lorentz and Hamiltonian gauges are in agreement.

\section{Semiclassical states}

In section 2 we considered "elementary" semiclassical  states  \r{2.5}
and their  superpositions \r{2.6} for scalar field models.  Let us now
write their analogs for the  scalar  electrodynamics  in  Hamiltonian,
Couloumb and Lorentz gauges and investigate their properties.

\subsection{Semiclassical states  in  refined  algebraic  quantization
approaches (Hamiltonian gauge)}

Consider state \r{2.5} in Hamiltonian gauge
\beq
\Psi_H[\varphi(\cdot)] = e^{\frac{i}{h}S}
e^{\frac{i}{h}\int d{\bf  x} \Pi({\bf x}) (\varphi({\bf x}) \sqrt{h} -
\Phi({\bf x}) )} g[\varphi(\cdot) - \frac{\Phi(\cdot)}{\sqrt{h}}]
\equiv K^h_X g[\varphi(\cdot)].
\l{4.1}
\eeq
For the simplicity, the following notations are introduced:
\bez
\varphi \equiv (A^k,\theta,\theta^*), \quad
\Phi \equiv ({\cal A}^k,\Theta,\Theta^*),
\quad
\Pi \equiv ({\cal E}_k,\Pi_{theta},\Pi_{\theta}^*);
\eez
$\varphi$ is field configuration,  $\Phi$ are classical fields,  $\Pi$
are classical momenta;  for integrals,  the followig simplification is
used:
\bez
\int d{\bf x} \Pi\Phi \equiv
\int d{\bf x}  [{\cal  E}_k  {\cal  A}^k  +  \Pi_{\theta}  \Theta^*  +
\Pi_{\theta}^*\Theta];
\quad
\int d{\bf x} \Pi\varphi \equiv
\int d{\bf x}  [{\cal  E}_k  {A}^k  +  \Pi_{\theta}  \theta^*  +
\Pi_{\theta}^*\theta].
\eez
Introduce a  special  notation  for  a  gauge  rransformation  in  the
configuration space:
\bez
\nu_{\alpha}\varphi \equiv \nu_{\alpha} (A,\theta,\theta^*) =
(A +  \frac{1}{\sqrt{h}}   \partial   \alpha,   \theta   e^{-i\alpha},
\theta^*e^{i\alpha}).
\eez
Then, the inner product \r{3.9} will be written as
\beq
(\Psi_H,\Psi_H) = \int D\alpha D\varphi
e^{\frac{i}{\sqrt{h}} \int d{\bf x}
\Pi({\bf x}) (\nu_{\alpha}\varphi({\bf x}) - \varphi({\bf x}))}
g^*[\varphi - \frac{\Phi}{\sqrt{h}}]
g[\nu_{\alpha} \varphi - \frac{\Phi}{\sqrt{h}}]
\l{4.2}
\eeq
Here the integration measure
\bez
D\varphi \equiv DA D\theta^* D\theta.
\eez
Notice that  quantities  $g$  and $g^*$ entering to expression \r{4.2}
are not exponentially small only if
\bez
\varphi - \frac{\Phi}{\sqrt{h}} \sim O(1),
\quad
\nu_{\alpha} \varphi - \frac{\Phi}{\sqrt{h}} \sim O(1).
\eez
Therefore, $\nu_{\alpha}\varphi - \varphi \sim O(1)$  and  $\alpha
\sim \sqrt{h}$.   Only   functions  $\alpha$  of  such  order  give  a
significant contribution to the integral  \r{4.2}.  To  calculate  the
inner product as $h\to 0$, perform a substitution
\bez
\alpha = \sqrt{h}\beta,
\quad
\varphi - \frac{\Phi}{\sqrt{h}} = \phi \equiv
(a^k,\vartheta,\vartheta^*).
\eez
Then the expresiion \r{4.2} will be taken to the form
\beb
(\Psi_H,\Psi_H) = \int D\beta D\phi
e^{\frac{i}{\sqrt{h}}\int d{\bf x}
\Pi({\bf x})
(\nu_{\beta\sqrt{h}}
(\frac{\Phi}{\sqrt{h}} + \phi)({\bf x}) -
(\frac{\Phi}{\sqrt{h}} + \phi)({\bf x}))}
g^*[\phi]
g [\nu_{\beta\sqrt{h}}
(\frac{\Phi}{\sqrt{h}} + \phi) -
\frac{\Phi}{\sqrt{h}}]
\\
\l{4.3}
\eeb
Let us evaluate the expressions entering to the inner product \r{4.3}.
One has
\bey
\nu_{\beta\sqrt{h}}
(\frac{\Phi}{\sqrt{h}} + \phi)
=
\nu_{\beta\sqrt{h}}
\left(
\frac{{\cal A}}{\sqrt{h}} + a,
\frac{{\Theta}}{\sqrt{h}} + \vartheta,
\frac{{\Theta^*}}{\sqrt{h}} + \vartheta^*
\right)
\\
=
\left(
\frac{{\cal A}}{\sqrt{h}} + a + \partial \beta,
\left[ \frac{{\Theta}}{\sqrt{h}} + \vartheta \right]
e^{-i\beta\sqrt{h}},
\left[ \frac{{\Theta^*}}{\sqrt{h}} + \vartheta^* \right]
e^{i\beta\sqrt{h}}
\right)
\eey
Therefore,
\bez
\nu_{\beta\sqrt{h}}
(\frac{\Phi}{\sqrt{h}} + \phi) -
(\frac{\Phi}{\sqrt{h}} + \phi) =
\left(
\partial \beta,
\left[ \frac{{\Theta}}{\sqrt{h}} + \vartheta \right]
(e^{-i\beta\sqrt{h}}-1),
\left[ \frac{{\Theta^*}}{\sqrt{h}} + \vartheta^* \right]
(e^{i\beta\sqrt{h}}-1)
\right),
\eez
so that
\bey
\int d{\bf x} \Pi
[\nu_{\beta\sqrt{h}}
(\frac{\Phi}{\sqrt{h}} + \phi) -
(\frac{\Phi}{\sqrt{h}} + \phi)]
= \\
\int d{\bf x}
\left\{
{\cal E}_k \partial_k\beta +
\Pi_{\theta}^*
\left[ \frac{{\Theta}}{\sqrt{h}} + \vartheta \right]
(e^{-i\beta\sqrt{h}}-1)
+
\Pi_{\theta}
\left[ \frac{{\Theta}^*}{\sqrt{h}} + \vartheta^* \right]
(e^{i\beta\sqrt{h}}-1)
\right\}
\eey
Notice that  the integrand in \r{4.3} is a product of a slowly varying
and damping at the  infinity  functional  by  the  rapidly  oscillatng
exponent
\bez
\exp\left\{
- \frac{i}{\sqrt{h}}
\int d{\bf x} \beta({\bf x})
(\partial_k {\cal E}_k + i
(\Pi_{\theta}^* \Theta - \Pi_{\theta} \Theta^*))
\right\}.
\eez
Therefore, the integral will be exponentially small,  except  for  the
case
\beq
\Lambda_{\bf x} \equiv
\partial_k {\cal E}_k + i
(\Pi_{\theta}^* \Theta - \Pi_{\theta} \Theta^*)
= 0.
\l{4.4}
\eeq
An analogous  fact was discovered in section 2:  it was found that the
"composed state"  \r{2.6}  is  exponentially  small  if   the   Maslov
isotropic condition is not satisfied.  Now we see that for constrained
systems additional  conditions  arise  even  for  wave  packet  states
\r{4.1}.

Under condition \r{4.4}, one can simplify expression
\r{4.3} as $h\to 0$. One should write
\bey
g
\left[\nu_{\beta\sqrt{h}}
(\frac{\Phi}{\sqrt{h}} + \phi) -
(\frac{\Phi}{\sqrt{h}} + \phi)\right] \simeq
g(a+ \partial\beta,  \vartheta - i\beta \Theta,  \vartheta^* +  i\beta
\Theta^*) =
\\
e^{\int d{\bf x} [\partial\beta \frac{\delta}{\delta a}
- i\beta \Theta \frac{\delta}{\delta \vartheta} +
i\beta \Theta^* \frac{\delta}{\delta \vartheta^*}] }
g(a,\vartheta,\vartheta^*),
\eey
use the Baker-Hausdorff formula for exponents and obtain that
\beq
(\Psi_H,\Psi_H) = \int D\phi
g^*[\phi] \int D\beta
e^{-i\int d{\bf x} \beta({\bf x}) (\Xi\Lambda_{\bf x})} g[\phi] =
(g,\prod_{\bf x} \delta(\Xi\Lambda_{\bf x}) g),
\l{4.5}
\eeq
where
\beq
\Xi\Lambda_{\bf x} \equiv
-i \partial_k \frac{\delta}{\delta a^k} +
i\Pi_{\theta}^* \vartheta + \Theta \frac{\delta}{\delta \vartheta}
- i \Pi_{\theta} \vartheta^* - \Theta^*
\frac{\delta}{\delta \vartheta^*}
\l{4.6}
\eeq
be a linearized constraint \r{4.4}. Since
\bez
[ \Xi \Lambda_{\bf x}, \Xi \Lambda_{\bf y}] = 0,
\eez
there are no operator ordering problems in \r{4.5}.

Thus, there  are  the  following   new   features   of   semiclassical
electrodynamics due to gauge symmetry.

First, not any classical configuration $X=(S,\Pi,\Phi)$ can be chosen:
the classical  constraint  condition  \r{4.4}  should  be   satisfied;
otherwise, state $K_X^hf$ will have zero norm.  Therefore, the base of
the semiclassical bundle (classical state space) is a "curved"
constraint surface in the flat space.

Next, the inner products in fibres ${\cal F}_X$ is $X$-dependent since
the linearized constraints $\Xi\Lambda_{\bf x}$ \r{4.6} depend on $X$.
The degenerate  inner product \r{4.5} resembles \r{2.16a}.  One should
consider usual procedures of factorization and completefication  of  a
pre-Hilbert space with inner product \r{4.5}.

The 1-forms $\omega$ and $\Omega$ have standard forms:
\beb
\omega_X[\delta X] = \int d{\bf x} \Pi({\bf x}) \delta \Phi({\bf x}) -
\delta S; \\
\Omega_X[\delta X] = \int d{\bf x}
\left[
\delta \Pi({\bf x}) \phi({\bf x}) - \delta \Phi({\bf x})
\frac{1}{i} \frac{\delta}{\delta \phi({\bf x})}
\right].
\l{4.6a}
\eeb
The commutation relations \r{2.14} are satisfied.

It is necessary  to  check  that  the  operator  $\Omega_X[\delta  X]$
conserves the equivalence property; it should take zero-norm states to
zero-norm states. To justify this property, one should prove that
\beq
[\Xi\Lambda_{\bf x};\Omega_X[\delta X]] = 0.
\l{4.6b}
\eeq
Notice that
\bez
\int d{\bf x} \beta({\bf x}) \Xi\Lambda_{\bf x} \equiv
- \Xi\Lambda[\beta] = - \Omega_X[\nabla_{\Lambda[\beta]}X],
\eez
where the infinitesimal vector
$
\nabla_{\Lambda[\beta]}X \equiv
(\nabla_{\Lambda[\beta]}S = 0,
\nabla_{\Lambda[\beta]}\Pi, \nabla_{\Lambda[\beta]}\Phi)$ has the form
of infinitesimal gauge transformation:
\beb
\nabla_{\Lambda[\beta]} {\cal E}_k = 0,
\quad
\nabla_{\Lambda[\beta]} \Pi_{\theta} = i\beta \Pi_{\theta},
\quad
\nabla_{\Lambda[\beta]} \Pi_{\theta}^* = -i\beta \Pi_{\theta}^*,
\\
\nabla_{\Lambda[\beta]} {\cal A}^k = -\partial_k\beta;
\quad
\nabla_{\Lambda[\beta]} \Theta = i\beta \Theta,
\quad
\nabla_{\Lambda[\beta]} \Theta^* = -i\beta \Theta^*
\l{4.6c}
\eeb
It follows from eq.\r{2.14x} and formula for $d\omega_X$ that
\bez
[\Xi\Lambda[\beta];\Omega_X[\delta X]] =
id\omega_X(\nabla_{\Lambda[\beta]}X;\delta X) =
- id\Lambda[\beta](\delta X)
\eez
with
\beq
\Lambda[\beta] \equiv \int d{\bf x} \beta({\bf x}) \Lambda_{\bf x}.
\l{4.6x}
\eeq
Notice that  $X$  and  $X+\delta X$ should both satisfy the additional
condition \r{4.4}  $\Lambda[\beta]=0$;  therefore,  quantity  \r{4.6x}
vanishes.

Let us investigate properties of the 1-form $\Omega$.  First  of  all,
notice that
\beq
\Omega_X [\nabla_{\Lambda[\beta]}X] g
= - \int d{\bf x} \beta({\bf x}) \Xi \Lambda_{\bf x} g \sim 0
\l{4.6d}
\eeq
for all  $g$  since state \r{4.6d} is orthogonal to all states because
of relation $\Xi\Lambda_{\bf x} \prod_{\bf  y}  \delta(\Xi\Lambda_{\bf
y}) = 0$. It also happens that the inverse statement is also valid:
\beq
\Omega_X[\delta X] \sim 0 \quad
\Rightarrow \quad
\delta X = (\delta S, \delta \Pi = \nabla_{\Lambda[\beta]}\Pi,
\delta \Phi = \nabla_{\Lambda[\beta]}\Phi).
\l{4.6e}
\eeq
To check implication \r{4.6e}, notice that
$\Omega_X[\delta X] \sim 0$ implies that
$[\Omega_X[\delta X];  \Omega_X[\delta X']] = 0$ for all infinitesimal
$\delta X'$, i.e.
\beb
d\omega_X(\delta X,\delta X') =
\int d{\bf x}
(\delta {\cal E}_k \delta {\cal A}^k{}' +
\delta  \Pi_{\theta}  \delta \Theta^*{}' +
\delta  \Pi^*_{\theta}  \delta \Theta'
- \delta {\cal A}^k \delta {\cal E}_k' -
\delta \Theta \delta \Pi_{\theta}^*{}' -
\delta \Theta^* \delta \Pi_{\theta}') = 0
\\
\l{4.6g}
\eeb
for $\delta X'$ satisfying the constraint relation
\beq
\partial_k \delta {\cal E}_k' +
i\delta(\Pi_{\theta}^*{}' - \Pi_{\theta}'\Theta^*{}') = 0.
\l{4.6h}
\eeq
Relation \r{4.6g} implies that $\delta {\cal E}_k = 0$,  $\delta {\cal
A}^k_{\perp} =  0$  since  $\delta  {\cal  A}^k{}'$  and $\delta {\cal
E}_{k\perp}'$ may be arbitrary.  Therefore,  $\delta {\cal  A}^k  =  -
\partial_k\beta$; making use of \r{4.6h}, we check statement \r{4.6e}.

For the following sections, it will be necessary to solve the equation
\beq
d\omega_X(\delta X,\delta X') = \kappa(\delta X'), \quad \delta X -?
\l{4.6i}
\eeq
where $\kappa$ is a given 1-form. It happens that problem \r{4.6i} has
a solution iff
\beq
\kappa(\nabla_{\Lambda[\beta]}X) = 0.
\l{4.6j}
\eeq
Namely, implication \r{4.6i}  $\to$  \r{4.6j}  is  evident.  To  check
implication \r{4.6j}  $\to$  \r{4.6i},  notice  that a general form of
$\kappa$ can be written as
\bez
\kappa(\delta X') =
\int d{\bf x}
(\delta {\cal E}_k \delta {\cal A}^k{}' +
\delta  \Pi_{\theta}  \delta \Theta^*{}' +
\delta  \Pi^*_{\theta}  \delta \Theta'
- \delta {\cal A}^k \delta {\cal E}_k' -
\delta \Theta \delta \Pi_{\theta}^*{}' -
\delta \Theta^* \delta \Pi_{\theta}') = 0
\eez
For $\delta    X    =    (\delta    S,    \delta   {\cal   E}_k,\delta
\Pi_{\theta},\delta \Pi_{\theta}^*,    \delta    {\cal     A}^k,\delta
\Theta,\delta \Theta^*)$,        eq.\r{4.6j}        implies       that
$d\Lambda[\beta](\delta X) = 0$.  Thus,  \r{4.6i}  $\equiv$  \r{4.6j}.
Note that the solution of problem \r{4.6i} is not unique:  one can add
to $\delta    X$    vector    $(\delta     S,
\delta     \Pi     = \nabla_{\Lambda[\beta]}\Pi,
\delta     \Phi     = \nabla_{\Lambda[\beta]}\Phi)$.

\subsection{Dirac semiclassical states. Couloumb gauge}

Let us rewrite state \r{4.1} in the  Dirac  approach.  Making  use  of
relation \r{3.i4}, one finds:
\beq
\Psi_D[\varphi(\cdot)] = \int D\alpha
e^{\frac{i}{h}S} e^{\frac{i}{\sqrt{h}} \int d{\bf x}
\Pi (\nu_{\alpha}\varphi - \frac{\Phi}{\sqrt{h}}) }
g[\nu_{\alpha}\varphi - \frac{\Phi}{\sqrt{h}}].
\l{4.7}
\eeq
Integral \r{4.7}  is  not  exponentially  small  only  if
$\nu_{\overline{\alpha}}\varphi  - \Phi/\sqrt{h} \sim O(1)$
for   some $\overline{\alpha}$, i.e.
\beq
a^k = A^k - \frac{{\cal A}^k -  \partial_k\overline{\alpha}}{\sqrt{h}}
\sim O(1), \quad
\vartheta = \theta e^{-i\overline{\alpha}}  -  \frac{\Theta}{\sqrt{h}}
\sim O(1).
\l{4.8}
\eeq
Denote $\phi  \equiv  (a^k,\vartheta,\vartheta^*)$   and   perform   a
substitution $\alpha = \overline{\alpha} + \sqrt{h}\beta$. One has:
\bez
\nu_{\overline{\alpha} + \sqrt{h}\beta} \varphi =
\left(
\frac{\cal A}{\sqrt{h}} + a + \partial\beta,
\left( \frac{\Theta}{\sqrt{h}} + \vartheta \right) e^{-i\sqrt{h}\beta},
\left( \frac{\Theta^*}{\sqrt{h}} + \vartheta^* \right)
e^{i\sqrt{h}\beta}
\right)
\eez
and
\bey
\int d{\bf x}
\Pi (\nu_{\alpha}\varphi - \frac{\Phi}{\sqrt{h}})
=
\\
\int d{\bf x}
[{\cal E}_k (a^k + \partial_k\beta)
+ \Pi_{\theta}^* \frac{\Theta}{\sqrt{h}}
(e^{-i\beta\sqrt{h}} - 1)
+ \Pi_{\theta}\frac{\Theta^*}{\sqrt{h}}
(e^{i\beta\sqrt{h}} - 1)
+ \Pi_{\theta}^* \vartheta e^{-i\beta\sqrt{h}}
+ \Pi_{\theta} \vartheta^* e^{i\beta\sqrt{h}}]
\eey
Under condition  \r{4.4},  one  finds  analogously  to  the   previous
subsection that
\beb
\Psi_D[\varphi(\cdot)] =
e^{\frac{i}{h}S}
e^{\frac{i}{\sqrt{h}} \int d{\bf x} \Pi({\bf x}) \phi({\bf x})}
f[\phi(\cdot)];
\\
f[\phi(\cdot)] =
\prod_{\bf x} \delta(\Xi\lambda_{\bf x}) g[\phi(\cdot)].
\l{4.9}
\eeb
For the Couloumb-gauge quantization,  we are interested only in values
of the Dirac functional $\Psi_D$ on the surface $\partial_k {\cal A}^k
= 0$.  Therefore,  the gauge function  $\overline{\alpha}$  should  be
chosen in    such    a    way   that   $\partial_k   ({\cal   A}^k   -
\partial_k\overline{\alpha}) = 0$.

Without loss of generality, one can specify classical states by sets
$X =  (S,{\cal  E}_k^{\perp},{\cal A}^k_{\perp},\Theta,\Pi_{\theta})$.
Then the semiclassical state \r{4.1} will be written as
\bey
\Psi_C[\varphi_{\perp}(\cdot)] =
e^{\frac{i}{h}S}
e^{\frac{i}{h}\int d{\bf x} \Pi_{\perp}({\bf x}) (\varphi_{\perp}({\bf
x}) \sqrt{h} - \Phi_{\perp}({\bf x})} f[\varphi_{\perp}(\cdot) -
\frac{\Phi_{\perp}(\cdot)}{\sqrt{h}}];
\eey
here
\bez
\varphi_{\perp} \equiv (A^k_{\perp},\theta,\theta^*),
\quad
\Phi_{\perp} \equiv ({\cal A}^k_{\perp},\Theta,\Theta^*),
\quad
\Pi_{\perp} \equiv
({\cal E}_k^{\perp},\Pi_{\theta},\Pi_{\theta}^*).
\eez
One has
\beq
(\Psi_C,\Psi_C) = \int D\phi_{\perp} |f[\phi_{\perp}]|^2
\l{4.10}
\eeq
for the inner product.  It is  possible  to  check  that  eqs.\r{4.5},
\r{4.9} and \r{4.10} indeed do not contradict each other.

The 1-forms are written analogously to \r{4.6a}:
\bey
\omega_X[\delta X]   =  \int  d{\bf  x}  \Pi_{\perp}({\bf  x})  \delta
\Phi_{\perp}({\bf x}) - \delta S; \\
\Omega_X[\delta X]  =  \int  d{\bf  x}  [\delta  \Pi_{\perp}({\bf  x})
\phi_{\perp}({\bf x}) -
\delta \Phi_{\perp}({\bf    x})    \frac{1}{i}    \frac{\delta}{\delta
\phi_{\perp}({\bf x})}.
\eey

\subsection{Semiclassical states   in   the   Gupta-Bleuler   approach
(Lorentz gauge)}

Consider the semiclassical states in the Gupta-Bleuler  approach.  Let
${\cal A}^0({\bf x})$ be some field configuration. Under condition
\bez
A^0({\bf x}) - \frac{{\cal A}^0({\bf x})}{\sqrt{h}} \equiv
a^0({\bf x}) \sim O(1),
\eez
suppose the Gupta-Bleuler state $\Psi_L$ be as follows
\bez
\Psi_L[\varphi_L(\cdot)] =
e^{\frac{i}{h}S} e^{\frac{i}{h}
\int d{\bf  x}  \Pi_L({\bf  x})   (\varphi_L({\bf   x})   \sqrt{h}   -
\Phi_L({\bf x}))} v[\varphi_L - \frac{\Phi_L}{\sqrt{h}}].
\eez
Here
\bez
\varphi_L \equiv (A^0,A^k,\theta,\theta^*),
\quad
\Phi_{L} = ({\cal A}^0,{\cal A}^k,\Theta,\Theta^*),
\quad
\Pi_{L} = ({\cal E}_0,{\cal E}_k,\Pi_{\theta},\Pi_{\theta}^*).
\eez
The values  of  $\Psi_L$ for arbitrary $A^0$ can be reconstructed from
the Gupta-Bleuler  condition  \r{3.i5}.  The  inner  product  has  the
following form:
\beq
(\Psi_L,\Psi_L) = \int D\lambda Da^k D\vartheta
D\vartheta^*
(v(-i\lambda,a,\vartheta,\vartheta^*))^*
v(i\lambda,a,\vartheta,\vartheta^*).
\l{4.11}
\eeq

Condition \r{3.i5}  implies property \r{4.4},  relation ${\cal E}_0=0$
and
\beq
\left(
\frac{1}{i} \frac{\delta}{\delta a^0({\bf x})}
- \frac{i}{\sqrt{-\Delta}} \Xi \Lambda_{\bf x}
\right) v(a^0,a^k,\vartheta,\vartheta^*) = 0.
\l{4.12}
\eeq
Consider the equivalence transformation \r{3.i5a} in the semiclassical
theory. For $Y_L = K^h_X\zeta$, one finds that
\bez
v \sim v +
\int d{\bf x}
\beta({\bf x})
\left(
\frac{1}{i} \frac{\delta}{\delta a^0({\bf x})}
+ \frac{i}{\sqrt{-\Delta}} \Xi \Lambda_{\bf x}
\right) \zeta.
\eez

It follows from eq.\r{4.12} that
\bez
v(a^0,a^k,\vartheta,\vartheta^*) =
e^{-\int d{\bf x} a^0({\bf x})
\frac{1}{\sqrt{-\Delta}} \Xi\Lambda_{\bf x} }
g[a^k,\vartheta,\vartheta^*]
\eez
with
\bez
g[a^k,\vartheta,\vartheta^*] =
v[0,a^k,\vartheta,\vartheta^*].
\eez
The inner product \r{4.11} is then in agreement with \r{4.5}.

The 1-forms $\omega$ and $\Omega$ have the standard forms:
\bey
\omega_X[\delta X] = \int d{\bf x} \Pi_L({\bf x})  \delta  \Phi_L({\bf
x}) - \delta S;
\\
\Omega_X[\delta X] = \int d{\bf x}
\left[\delta \Pi_L({\bf x}) \phi_L({\bf x}) -
\delta \Phi_L({\bf  x})  \frac{1}{i}  \frac{\delta}{\delta \phi_L({\bf
x})}\right]
\eey
One should check that the operators $\Omega_X[\delta X]$ conserve  the
additional condition \r{4.12}. It is sufficient to justify that
\bez
\left[
\frac{1}{i} \frac{\delta}{\delta a^0({\bf x})} ;
\Omega_X[\delta X]
\right] = 0,
\quad
[\Xi\Lambda[\beta];\Omega_X[\delta X]] = 0.
\eez
The first property means that $\delta {\cal E}_0 = 0$,  the second  is
checked  analogously to \r{4.6b}.

\subsection{Comparison of different gauges}

Let us  compare semiclassical electrodynamics in different gauges (see
table 1).

\begin{table}
\caption{
Semiclassical bundle for different gauges
}
{\footnotesize
\begin{tabular}{|p{2cm}|p{14cm}|}
\hline
Base $\cal X$ of the semiclassical bundle
&
\begin{description}
\item[Hamiltonian gauge:]
Set of all $(S,{\cal E}_k({\bf x}),{\cal A}^k({\bf x}),
\Pi_{\theta}({\bf x}),\Theta({\bf x}))$ such that
$\partial_k{\cal E}_k + i(\Pi_{\theta}^*\Theta - \Pi_{\theta}
\Theta^*)=0$
\item[Couloumb gauge:]
Set of all $(S,{\cal E}^{\perp}_k({\bf x}),
{\cal A}^k_{\perp}({\bf x}))$
\item[Lorentz gauge:]
Set of all $(S,{\cal E}_0({\bf x}),{\cal E}_k({\bf x}),
{\cal A}^0({\bf x}),{\cal A}^k({\bf x}),
\Pi_{\theta}({\bf x}),\Theta({\bf x}))$ such that
${\cal E}_0=0$ and
$\partial_k{\cal E}_k + i(\Pi_{\theta}^*\Theta - \Pi_{\theta}
\Theta^*)=0$
\end{description}
\\ \hline
A fibre ${\cal F}_X$, $X\in {\cal X}$
&
\begin{description}
\item[Hamiltonian gauge:]
Space of functionals $g[a^k,\vartheta,\vartheta^*]$ with inner product
$\int Da^k  D\vartheta   D\vartheta^*   g^*   \prod_{\bf   x}   \delta
(\Xi\Lambda_{\bf x}) g$. The space should be factorized and completed.
\item[Couloumb gauge:]
Space of functionals $f[a^k_{\perp},\vartheta,\vartheta^*]$
with inner product
$\int Da^k_{\perp} D\vartheta D\vartheta^* f^* f$.
\item[Lorentz gauge:]
Space of functionals
$v[a^0,a^k,\vartheta,\vartheta^*]$ with inner product
$\int D\lambda Da^k D\vartheta
D\vartheta^*
(v(-i\lambda,a,\vartheta,\vartheta^*))^*
v(i\lambda,a,\vartheta,\vartheta^*)$. An additional condition
$\left(
\frac{1}{i} \frac{\delta}{\delta a^0({\bf x})}
- \frac{i}{\sqrt{-\Delta}} \Xi \Lambda_{\bf x}
\right) v(a^0,a^k,\vartheta,\vartheta^*) = 0$ is imposed;
states
$v \sim v +
\int d{\bf x}
\left(
\frac{1}{i} \frac{\delta}{\delta a^0({\bf x})}
+ \frac{i}{\sqrt{-\Delta}} \Xi \Lambda_{\bf x}
\right) \zeta$ are set to be equivalent.
\end{description}
\\ \hline
Correspon\-dence between gauges
&
\begin{description}
\item[Hamiltonian and Couloumb gauges:]
$f = \prod_{\bf x} \delta(\Xi\Lambda_{\bf x}) g$
\item[Hamiltonian and Lorentz gauges:]
$v(a^0,a^k,\vartheta,\vartheta^*) =
e^{-\int d{\bf x} a^0({\bf x})
\frac{1}{\sqrt{-\Delta}} \Xi\Lambda_{\bf x} }
g[a^k,\vartheta,\vartheta^*]$
\end{description}
\\ \hline
1-form $\omega$
&
\begin{description}
\item[Hamiltonian gauge:]
$\omega_X[\delta X] = \int d{\bf x} [{\cal E}_k \delta {\cal A}^k  +
\Pi_{\theta} \delta \Theta^* + \Pi_{\theta}^* \delta \Theta] - \delta S$
\item[Couloumb gauge:]
$\omega_X[\delta X] = \int d{\bf x} [{\cal E}_k^{\perp}
\delta {\cal A}^k_{\perp}  +
\Pi_{\theta} \delta \Theta^* + \Pi_{\theta}^* \delta \Theta] - \delta S$
\item[Lorentz gauge:]
$\omega_X[\delta X] = \int d{\bf x} [{\cal E}_k \delta {\cal A}^k  +
\Pi_{\theta} \delta \Theta^* + \Pi_{\theta}^* \delta \Theta] - \delta S$
\end{description}
\\ \hline
1-form $\Omega$
&
\begin{description}
\item[Hamiltonian gauge:]
$\Omega_X[\delta X] = \int d{\bf x} [\delta {\cal E}_k
a^k + \delta \Pi_{\theta} \vartheta^* + \delta \Pi_{\theta}^* \vartheta
- \delta {\cal A}^k \frac{1}{i} \frac{\delta}{\delta a^k}
- \delta \Theta \frac{1}{i} \frac{\delta}{\delta \vartheta}
- \delta \Theta^* \frac{1}{i} \frac{\delta}{\delta \vartheta^*}]$
\item[Couloumb gauge:]
$\Omega_X[\delta X] = \int d{\bf x} [\delta {\cal E}_k^{\perp}
a^k_{\perp}
+ \delta \Pi_{\theta} \vartheta^* + \delta \Pi_{\theta}^* \vartheta
- \delta   {\cal   A}^k_{\perp}    \frac{1}{i}
\frac{\delta}{\delta a^k_{\perp}}
- \delta \Theta \frac{1}{i} \frac{\delta}{\delta \vartheta}
- \delta \Theta^* \frac{1}{i} \frac{\delta}{\delta \vartheta^*}]$
\item[Lorentz gauge:]
$\Omega_X[\delta X] = \int d{\bf x} [\delta {\cal E}_k
a^k + \delta \Pi_{\theta} \vartheta^* + \delta \Pi_{\theta}^* \vartheta
- \delta {\cal A}^0 \frac{1}{i} \frac{\delta}{\delta a^0}
- \delta {\cal A}^k \frac{1}{i} \frac{\delta}{\delta a^k}
- \delta \Theta \frac{1}{i} \frac{\delta}{\delta \vartheta}
- \delta \Theta^* \frac{1}{i} \frac{\delta}{\delta \vartheta^*}]$
\end{description}
\\ \hline
Zero operators $\Omega_X[\delta X]\sim 0$
&
\begin{description}
\item[all gauges:]
$\delta X = (\delta S,\delta \Pi=0,\delta \Phi=0)$
\item[Hamiltonian and Lorentz gauges:]
$\delta X = \nabla_{\Lambda[\beta]}X$;
this means that
$\nabla_{\Lambda[\beta]} {\cal E}_k = 0$,
$\nabla_{\Lambda[\beta]} \Pi_{\theta} = i\beta \Pi_{\theta}$,
$\nabla_{\Lambda[\beta]} \Pi_{\theta}^* = -i\beta \Pi_{\theta}^*$,
$\nabla_{\Lambda[\beta]} {\cal A}^k = -\partial_k\beta$;
$\nabla_{\Lambda[\beta]} \Theta = i\beta \Theta$,
$\nabla_{\Lambda[\beta]} \Theta^* = -i\beta \Theta^*$
\item[Lorentz gauge:]
$\delta X = \nabla_{{\cal E}_0[\kappa]}X$:
this means that
$\nabla_{{\cal E}_0[\kappa]} {\cal A}^0 = \beta$,  other variations are
zero.
\end{description}
\\ \hline
\end{tabular}
}
\end{table}

An important  specific  feature of gauge theories is that the operator
1-form $\Omega_X[\delta X]$ may be zero for some $\delta X$ not of the
trivial form $\delta S\ne 0$, $\delta \Pi = 0$, $\delta \Phi = 0$.
All "zero modes" of the 1-form are presented in table 1.
Check of properties of zero operators $\Omega_X[\delta X]$ for the Lorentz
gauge case is analogous to the Hamiltonian gauge case.

Analogously to the scalar theories,  one can check that  any  operator
commuting with  all  $\Omega_X[\delta  X]$  is  a  multiplicator  by a
c-number constant.  For the Couloumb gauge,  with no gauge freedom, it
is evident; other gauges are equivalent to it.

\section{Semiclassical observables and transformations}

Field operators   are  important  objects  of  quantum  field  theory.
However, for     gauge     theories     fields     $\hat{A}^{\mu}(x)$,
$\hat{\theta}(x)$ are  not physical observables.  Therefore,  the more
complicated gauge-invariant combinations of fields should be viewed as
observables.

In classical  mechanics,  observables  may  be introduced in different
ways. First,  one can say  that  states  of  a  classical  system  are
specified by  points  of the phase space and an observable is given if
its value in any state is specified.  Observables then are  viewed  as
real functions on the phase space.

Alternatively, any  observable  may  be  also  viewed  as  a classical
Hamiltonian generating an evolution transformation  group.  Thus,  one
can say  that  an observable is specified if a one-parametric group of
symplectic transformation is given.

Analogously, quantum  observables  may  be  specified   by   Hermitian
operators, as well as by unitary evolution groups.

These conceptions  may  be  considered  in the semiclassical theory as
well. Subsection  5.1  deals  with  semiclassical   investigation   of
observables viewed  as Hermitian operators.  Subsection 5.2 is devoted
to semiclassical analysis  of  evolution  generated  by  semiclassical
observables. Examples   are   gauge   and   Poincare   transformations
(subsections 5.3 and 5.4).

Heisenberg fields are very important objects of quantum field  theory.
Their semiclassical analogs are investigated in subsection 5.5.

Gauge equivalence   relation   should    conserve    under    Poincare
transformations: gauge equivalent semiclassical states should be taken
to gauge equivalent. This property is discussed in subsection 5.6.

To check Poincare group relation \r{2.26},  it is convenient to reduce
it to its infinitesimal Lie algebra analog. This problem is considered
in subsection 5.7.

\subsection{Semiclassical observables}

First of all consider the observables  in  the  Hamiltonian  approach.
Suppose them  to  depend  on  the  small parameter $\sqrt{h}$,  fields
$\hat{\varphi} =   (\hat{A}^{\mu},\hat{\theta},\hat{\theta}^*)$    and
momenta $\hat{\pi} =
(\hat{E}_{\mu},\hat{\pi}_{\theta},\hat{\pi}_{\theta}^*)$ as
\beq
\hat{O}^h = O(\sqrt{h}\hat{\varphi}, \sqrt{h}\hat{\pi}).
\l{5.1}
\eeq
Operators \r{5.1}  will  be  called  semiclassical.  It is supposed in
qunatum field theory  that  expression  \r{5.1}  is  well-defined  iff
$O(\Phi,\Pi)$ is  a  gauge-invariant functional:  it should not change
under transformation
\bez
\Theta \to \Theta e^{-i\alpha},
\quad
\Pi_{\theta} \to \Pi_{\theta} e^{-i\alpha},
\quad
{\cal A}^k \to {\cal A}^k + \partial_k \alpha,
\quad
{\cal A}^0 \to {\cal A}^0 + \kappa.
\eez
Apply the semiclassical operator \r{5.1} to  the  semiclassical  state
$K^h_Xf$. The general structure of the commutation rule is aas follows
\beq
\hat{O}^h K^h_Xf =
K^h_X (O(X) + \sqrt{h} (\Xi O)(X) + \frac{h}{2}
(\Xi^2O)(X) + ...) f.
\l{5.2}
\eeq
For different gauges,  explicit forms of the operators  $\Xi^nO$  (for
example, $\Xi O$) are presented in table 2.

\begin{table}
\caption{
Semiclassical observables for different gauges
}
{\footnotesize
\begin{tabular}{|p{2cm}|p{14cm}|}
\hline
Operators $\Xi^nO(X)$
&
\begin{description}
\item[Hamiltonian gauge:]
\bey
(\Xi^n_H O)(X) = \frac{\partial^n}{\partial \sqrt{h}^n}|_{h=0}
O({\cal A}^k + \sqrt{h}a^k,  {\cal E}_k  +  
\frac{\sqrt{h}}{i}  \frac{\delta}{\delta
a^k},
\\
\Theta   +  \sqrt{h}\vartheta,  \Theta^*  +  \sqrt{h}\vartheta^*,
\Pi_{\theta}  +  \frac{\sqrt{h}}{i} \frac{\delta}{\delta \vartheta^*},
\Pi_{\theta}^*  +  \frac{\sqrt{h}}{i} \frac{\delta}{\delta \vartheta})
\eey
\item[Couloumb gauge:]
\bey
(\Xi^n_C O)(X) = \frac{\partial^n}{\partial \sqrt{h}^n}|_{h=0}
O({\cal A}^k_{\perp} + \sqrt{h}a^k_{\perp},  {\cal E}_k  +
\sqrt{h} \hat{\varepsilon}_k     -
h \frac{i}{\partial^2}     \partial_k
({\vartheta} \frac{1}{i} \frac{\delta}{\delta \vartheta}  - {\varphi}^*
\frac{1}{i} \frac{\delta}{\delta \vartheta^*}),
\\
\Theta   +  \sqrt{h}\vartheta,  \Theta^*  +  \sqrt{h}\vartheta^*,
\Pi_{\theta}  +  \frac{\sqrt{h}}{i} \frac{\delta}{\delta \vartheta^*},
\Pi_{\theta}^*  +  \frac{\sqrt{h}}{i} \frac{\delta}{\delta \vartheta}),
\\
{\cal E}_k  =  (\delta_{kl} - \frac{\partial_k\partial_l}{\partial^2})
{\cal E}_k^{\perp} - \frac{1}{\partial^2} \partial_k  i  (\Pi_{\theta}^*\Theta  -
\Pi_{\theta} \Theta^*);
\\
\hat{\varepsilon}_k =                  (\delta_{kl}                  -
\frac{\partial_k\partial_l}{\partial^2})                   \frac{1}{i}
\frac{\delta}{\delta a_{\perp}^l}  -  \frac{1}{\partial^2}  \partial_k
(i\Pi_{\theta}^*\vartheta + \Theta \frac{\delta}{\delta \vartheta}
- i\Pi_{\theta} \vartheta^* - \Theta^* \frac{\delta}{\delta \vartheta^*})
\eey
\item[Lorentz gauge:]
\bey
(\Xi^n_L O)(X) = \frac{\partial^n}{\partial \sqrt{h}^n}|_{h=0}
O({\cal A}^k + \sqrt{h}a^k,
{\cal E}_k  +  \frac{\sqrt{h}}{i}  \frac{\delta}{\delta
a^k}, {\cal A}^0 + \sqrt{h}a^0,
\frac{\sqrt{h}}{i} \frac{\delta}{\delta a^0},
\\
\Theta   +  \sqrt{h}\vartheta,  \Theta^*  +  \sqrt{h}\vartheta^*,
\Pi_{\theta}  +  \frac{\sqrt{h}}{i} \frac{\delta}{\delta \vartheta^*},
\Pi_{\theta}^*  +  \frac{\sqrt{h}}{i} \frac{\delta}{\delta \vartheta}).
\eey
\end{description}
\\ \hline
Partial case: $\Xi O(X)$
&
\begin{description}
\item[Hamiltonian gauge:]
$\Xi_H O = \int d{\bf x} (
\frac{\delta O}{\delta  {\cal  A}^k({\bf x})} a^k({\bf x})
+ \frac{\delta O}{\delta
{\cal E}_k({\bf x})} \frac{1}{i} \frac{\delta}{\delta a^k({\bf x})}
+ \frac{\delta O}{\delta \Theta({\bf x})} \vartheta({\bf x}) +
\frac{\delta O}{\delta \Theta^*({\bf x})} \vartheta^*({\bf x}) +
\frac{\delta O}{\delta \Pi_{\theta}({\bf x})} \frac{1}{i}
\frac{\delta}{\delta  \vartheta^*({\bf x})} +
\frac{\delta O}{\delta \Pi_{\theta}^*({\bf x})} \frac{1}{i}
\frac{\delta}{\delta  \vartheta({\bf x})}
)$
\item[Couloumb gauge:]
$\Xi_C O = \Xi_H O - \int d{\bf x} \left(
\frac{\delta O}{\delta   {\cal   A}^k({\bf  x})}  \frac{1}{\partial^2}
\partial_k \Xi_H \partial_l {\cal A}^l({\bf x}) +
\frac{\delta O}{\delta   {\cal   E}_k({\bf  x})}  \frac{1}{\partial^2}
\partial_k \Xi_H {\Lambda}_{\bf x}
\right)$
\item[Lorentz gauge:]
$
\Xi_L O = \Xi_H O + \int d{\bf x} \left(
\frac{\delta O}{\delta  {\cal  A}^0({\bf x})} a^0({\bf x})
+ \frac{\delta O}{\delta
{\cal E}_0({\bf x})} \frac{1}{i} \frac{\delta}{\delta a^0({\bf x})}
\right)
$
\end{description}
\\ \hline
$\Xi O(X) = - \Omega[\nabla_OX]$; form of $\nabla_OX$
&
\begin{description}
\item[Hamiltonian gauge:]
$\nabla_O {\cal A}^k = \frac{\delta O}{\delta {\cal E}_k}$;
$\nabla_O {\cal E}_k = - \frac{\delta O}{\delta {\cal A}^k}$;
$\nabla_O{\Theta} = \frac{\delta O}{\delta \Pi_{\theta}^*}$;
$\nabla_O {\Pi_{\theta}} = - \frac{\delta O}{\delta \Theta^*}$;
$\nabla_O {\Theta^*} = \frac{\delta O}{\delta \Pi_{\theta}}$;
$\nabla_O{\Pi_{\theta}^*} = - \frac{\delta O}{\delta \Theta}$.
\item[Couloumb gauge:]
$\nabla_O {\cal A}^k =  \frac{\delta  O}{\delta  {\cal  E}_k}  -  \partial_k
\frac{1}{\partial^2} \partial_l \frac{\delta O}{\delta {\cal E}_l}$;
$\nabla_O {\cal E}_k^{\perp} = - \frac{\delta O}{\delta {\cal A}^k}$;
$\nabla_O{\Theta} = \frac{\delta O}{\delta \Pi_{\theta}^*} + i \Theta
\frac{1}{\partial^2} \partial_l \frac{\delta O}{\delta {\cal E}_l}$;
$\nabla_O{\Pi_{\theta}} = - \frac{\delta O}{\delta \Theta^*}
+ i\Pi_{\theta}
\frac{1}{\partial^2} \partial_l \frac{\delta O}{\delta {\cal E}_l}$;
$\nabla_O{\Theta^*} = \frac{\delta O}{\delta \Pi_{\theta}} - i \Theta^*
\frac{1}{\partial^2} \partial_l \frac{\delta O}{\delta {\cal E}_l}$;
$\nabla_O{\Pi_{\theta}^*} = - \frac{\delta O}{\delta \Theta} - i\Pi_{\theta}^*
\frac{1}{\partial^2} \partial_l \frac{\delta O}{\delta {\cal E}_l}$.
\item[Lorentz gauge:]
$\nabla_O{\cal A}^k = \frac{\delta O}{\delta {\cal E}_k}$;
$\nabla_O{\cal E}_k = - \frac{\delta O}{\delta {\cal A}^k}$;
$\nabla_O{\cal A}^0 = \frac{\delta O}{\delta {\cal E}_0}$;
$\nabla_O{\cal E}_0 = - \frac{\delta O}{\delta {\cal A}^0}$;
$\nabla_O{\Theta} = \frac{\delta O}{\delta \Pi_{\theta}^*}$;
$\nabla_O{\Pi_{\theta}} = - \frac{\delta O}{\delta \Theta^*}$;
$\nabla_O{\Theta^*} = \frac{\delta O}{\delta \Pi_{\theta}}$;
$\nabla_O{\Pi_{\theta}^*} = - \frac{\delta O}{\delta \Theta}$.
\end{description}
\\ \hline
Gauge invariance condition
&
\begin{description}
\item[Hamiltonian and Lorentz gauges:]
$d\Lambda[\beta][\nabla_OX]=0$
\item[Lorentz gauge:]
$d{\cal E}_0[\beta][\nabla_OX]=0$
\end{description}
\\ \hline
\end{tabular}
}
\end{table}

One can take the operator $\Xi O$ to the canonical form \r{2.33}.  The
tangent vectors $\nabla_O X$ to the base of the  semiclassical  bundle
are also  calculated  in  table  2  for  different gauges ($\nabla_OX$
appears to be a tangent vector,  provided that  the  gauge  invariance
conditions are  satisfied).  Since the operaotrs $\Omega_X[\nabla_OX]$
are well-defined according to the previous section,  the operator $\Xi
O(X)$ also  takes  zero-norm  states to zero-norm states and conserves
the linearized Gupta-Bleuler condition \r{4.12}.

\subsection{Transformations of semiclassical states}

{\bf 1.} Poincare and gauge transformations are of the form
\beq
\exp \left[ - \frac{i}{h} \tau \hat{O} \right],
\l{5.3}
\eeq
where $\hat{O}$   is   of  the  semiclassical  form  \r{5.1}.  Due  to
renormalization, one  should  also  take  into  account  the  one-loop
quantum corrections and consider observables of the more general form
\bez
\hat{O} = O(\sqrt{h}\hat{\varphi},\sqrt{h}\hat{\pi}) +
h O^{(1)}(\sqrt{h}\hat{\varphi},\sqrt{h}\hat{\pi}) + ...
\eez
Let us consider the state vector
\bez
\Psi^{\tau} \equiv e^{-\frac{i}{h}\hat{O}\tau} K^h_{X_0} f_0
\eez
as $h\to 0$. It satisfies the following Cauchy problem
\beq
i \frac{d\Psi^{\tau}}{d\tau} = \frac{1}{h} \hat{O} \Psi^{\tau},
\quad \Psi^0 = K^h_{X_0} f_0.
\l{5.4}
\eeq
Let us look for the approximate as $h\to 0$ solutions  of  the  Cauchy
problem \r{5.4} in a following form:
\beq
\Psi^{\tau} \simeq K^h_{X^{\tau}} f^{\tau}.
\l{5.5}
\eeq
Let us find semiclassical equations  for  $X^{\tau}$,  $f^{\tau}$.  It
happens that commutation rule \r{2.12} is to be corrected for our case
as:
\beq
ih \frac{d}{d\tau} K^h_{X^{\tau}} f^{\tau} =
K^h_{X^{\tau}}
[\omega_{X^{\tau}} [\dot{X}^{\tau}]] -
\sqrt{h} \Omega_{X^{\tau}}[\dot{X}^{\tau}]
+ ih \frac{d}{d\tau}] f^{\tau};
\l{5.6}
\eeq
no additional   terms   of  the  order  $O(h)$  are  added.  Combining
commutation rules \r{5.5} and \r{5.6},  one  finds  that  substitution
\r{5.5} is an approximate solution of eq.\r{5.4} iff
\beq
\omega_{X^{\tau}}[\dot{X}^{\tau}] = O(X^{\tau});
\quad
\Omega_{X^{\tau}}[\dot{X}^{\tau}] = - \Xi O(X^{\tau});
\l{5.7}
\eeq
\beq
i \frac{d}{d\tau} f^{\tau} =
\left[
\frac{1}{2} \Xi^2 O((X^{\tau}) + O^{(1)}(X^{\tau})
\right] f^{\tau}.
\l{5.8}
\eeq
Eqs. \r{5.7} specify classical evolution. The first relation allows us
to express  $\dot{S}^{\tau}$  via  other   derivatives.   The   second
equations can be written as
\bez
\dot{X}^{\tau} = \nabla_O X^{\tau} + \overline{\delta}X,
\quad
\Omega_X[\overline{\delta}X] = 0.
\eez
We see  that $\dot{X}^{\tau}$ is defined up to a gauge transformation.
One can choose $\overline{\delta}X$ in order to  make  equations  more
convenient.

Denote by  $u^{\tau}_O:  X  \mapsto  u_O^{\tau}  X$ the transformation
taking the  initial  conditions  for  the  system  of  equations   for
$X^{\tau} \equiv (S^{\tau},\Pi^{\tau},\Phi^{\tau})$ of the form
\beq
\dot{S}^{\tau} =
\int d{\bf x} \Pi^{\tau} \dot{\Phi}^{\tau} -
O(\Pi^{\tau},\Phi^{\tau}),
\quad
\dot{\Pi}^{\tau} = \nabla_O\Pi^{\tau},
\quad
\dot{\Phi}^{\tau} = \nabla_O\Phi^{\tau}
\l{5.9}
\eeq
to the solution of the Cauchy problem. This is the classical evolution
corresponding to the observable $O$.

{\bf 2.}  Investigate  the  properties  of evolution of $f^t$ which is
given by eq.\r{5.8}.  Let $X^{\tau}(\alpha)$ be a function  of  $\tau$
and $\alpha = (\alpha_1,...,\alpha_k)$.  It happens that the following
relation is satisfied:
\beb
[i \frac{d}{d\tau} -
\frac{1}{2} \Xi^2O(X^{\tau}) - O^{(1)}(X^{\tau});
\Omega_{X^{\tau}}[\frac{\partial X^{\tau}}{\partial \alpha_a}]] =
\\
i \Omega_{X^{\tau}}
[\frac{\partial}{\partial \alpha_a}
(\frac{\partial X^{\tau}}{\partial \tau} -
\nabla_O X^{\tau})]
\l{5.10}
\eeb
One can check equality \r{5.10} in  differeny  ways.  First,  one  can
start from the identity
\bez
[ih\frac{d}{d\tau} - \hat{O}^h, ih\frac{\partial}{\partial \alpha_a}]
K^h_{X^{\tau}(\alpha)} f^{\tau}(\alpha) = 0.
\eez
It is taken to the following form
\bey
[\omega_X[\dot{X}] - \sqrt{h} \Omega_X[\dot{X}] +
ih \frac{\partial}{\partial \tau} -
O(X) - \sqrt{h} \Xi O(X) - \frac{h}{2} \Xi^2O(X) - O^{(1)}(X);
\\
\omega_X[\frac{\partial X}{\partial \alpha_a}
-\sqrt{h} \Omega_X[\frac{\partial X}{\partial \alpha_a}] +
ih \frac{\partial}{\partial \alpha_a}] = 0.
\eey
Considering the terms of the order  $O(h^{3/2})$,  one  comes  to  the
identity \r{5.10}.  Another  way  to  check  eq.\r{5.10} is to use the
direct calculation method. It is important to notice that the operator
identity \r{5.10}   is   valid  even  for  the  space  of  functionals
$g[a^k,\vartheta,\vartheta^*]$ before factorization.

{\bf 3. } Let us investigate the unitariry proprty for the Hamiltonian
gauge. It happens that one should require that
\beq
[\nabla_O;\nabla_{\Lambda[\beta]}] = - \nabla_{\Lambda[C_O\beta]}
\l{5.11}
\eeq
for some linear operator $C_O$. Under condition \r{5.11}, let us check
that equality
\beb
[i \frac{d}{d\tau} -
\frac{1}{2} \Xi^2O(X^{\tau}) - O^{(1)}(X^{\tau});
(\Xi\Lambda[\beta])(X^{\tau})] =
\\
i (\Xi\Lambda[C_O\beta])(X^{\tau})]
\l{5.12}
\eeb
is valid for the space of  functionals  $g[a^k,\vartheta,\vartheta^*]$
before factorization.

To justify property \r{5.12}, notice that it is equivalent to
\beb
[i \frac{d}{d\tau} -
\frac{1}{2} \Xi^2O(X^{\tau}) - O^{(1)}(X^{\tau});
\Omega_{X^{\tau}}[\nabla_{\Lambda[\beta]} X^{\tau}]
=
\\
i \Omega_{X^{\tau}}[\nabla_{\Lambda[C_O \beta]} X^{\tau}].
\l{5.13}
\eeb
To check relation \r{5.13}, leu us use identity \r{5.10}. Set
\bez
X^{\tau}(\alpha) \equiv  u^{\alpha}_{\Lambda[\beta]}  u^{\tau}_O  X  =
u^{\alpha}_{\Lambda[\beta]}X^{\tau}.
\eez
Then
$\nabla_{\Lambda[\beta]} X^{\tau}         \equiv        \frac{\partial
X^{\tau}(\alpha)}{\partial \alpha}$ and property \r{5.13} is taken to
\bey
\nabla_{\Lambda[C_O\beta]} X^{\tau} =
\frac{\partial}{\partial \alpha}\left|_{\alpha=0}
\left(
\frac{\partial X^{\tau}(\alpha)}{\partial \tau} -
\nabla_O X^{\tau}(\alpha)
\right)
\right. =
\\
\frac{\partial}{\partial \alpha}\left|_{\alpha=0}
\frac{\partial}{\partial t}\left|_{t=0}
\left(
u^{\alpha}_{\Lambda[\beta]} u^{t+\tau}_O -
u^t_O u^{\alpha}_{\Lambda[\beta]} u^{\tau}_O
\right) X
\right. \right. =
[\nabla_{\Lambda[\beta]};\nabla_O] X^{\tau}.
\eey
Thus, relation \r{5.12} is satisfied.

Let us now check conservation of the inner product
\bez
(g^{\tau},\prod_{\bf x} \delta(\Xi\Lambda_{\bf x}) (X^0) g^{\tau})
= (g^{\tau}, \int D\beta e^{i (\Xi\Lambda[\beta])(X^{\tau})} g^{\tau})
\eez
where
\bez
(g,\tilde{g}) \equiv Da^k D\vartheta D\vartheta^*
g^*[a^k,\vartheta,\vartheta^*]
\tilde{g}[a^k,\vartheta,\vartheta^*]
\eez
It follows from eq.\r{5.12} that
\bey
[i \frac{d}{d\tau} -
\frac{1}{2} \Xi^2O(X^{\tau}) - O^{(1)}(X^{\tau});
e^{i \Xi\Lambda[\beta] (X^{\tau}) } ]
\\ =
- \Xi \Lambda[C_O\beta](X^{\tau})
e^{i \Xi\Lambda[\beta] (X^{\tau})} =
- \int d{\bf y} (C_O\beta)({\bf y})
\frac{1}{i} \frac{\delta}{\delta \beta({\bf y})}
e^{i \Xi\Lambda[\beta] (X^{\tau}) } ].
\eey
Therefore,
\bez
i \frac{d}{d\tau}
(g^{\tau},\prod_{\bf x} \delta(\Xi\Lambda_{\bf x}) (X^0) g^{\tau})
=
[O^{(1)}(X^{\tau}) - O^{(1)*}(X^{\tau}) - i Tr C_O(X^{\tau})]
(g^{\tau},\prod_{\bf x} \delta(\Xi\Lambda_{\bf x}) (X^0) g^{\tau}).
\eez
Thus, zero-norm states are always  taken  to  zero-norm  states  under
condition \r{5.12}, while unitary requirements mean that
\beq
Im O^{(1)} = \frac{1}{2} Tr C_O.
\l{5.14}
\eeq
Usually, $Tr C_O$ will vanish.

{\bf 4.} For the Lorentz gauge,  one should check conservation of  the
linearized Gupta-Bleuler  condition.  A  sufficient  condition  is  as
follows:
\beb
[i \frac{d}{d\tau} -
\frac{1}{2} \Xi^2O(X^{\tau}) - O^{(1)}(X^{\tau});
\Xi{\cal E}_0[\beta] (X^{\tau})
- i \Xi\Lambda[\frac{1}{\sqrt{-\Delta}}\beta](X^{\tau})] =
\\
i (\Xi{\cal E}_0[C_O^{(L)} \beta] (X^{\tau})
- i \Xi\Lambda[\frac{1}{\sqrt{-\Delta}}C_O^{(L)}\beta](X^{\tau})).
\l{5.15}
\eeb
for some operator $C_O^{(L)}$.
It is a corollary of the relation
\beq
[\nabla_O;
\nabla_{{\cal E}_O[\beta] - i\Lambda [\frac{1}{\sqrt{-\Delta}}\beta]}]
=
\nabla_{{\cal E}_O[C_O^{(L)}\beta] -
i\Lambda [\frac{1}{\sqrt{-\Delta}}C_O^{(L)}\beta]}.
\l{5.16}
\eeq

{\bf 5.}   Thus,   for   all   observables  we  have  constructed  the
semiclassical evolution transformation taking  initial  condition  for
eq.\r{5.8} to  the  solution  for  this equation.  This transformation
conserves equivalence property and inner product. It can be reduced to
the factorspace; denote the obtained operator as
$U_O^{\tau}(u_O^{\tau}X\gets X):  {\cal F}_X \to {\cal F}_{u^{\tau}_OX}$.

The introduced transformations $u_O^{\tau}$ and
$U_O^{\tau}(u_O^{\tau}X\gets X)$ obey the  following  properties.  Let
$X=X(\alpha)$; then
\beb
\omega_{u_OX} [\frac{\partial (u_OX)}{\partial \alpha_a}] =
\omega_X [\frac{\partial X}{\partial \alpha_a}];
\\
\Omega_{u_OX} [\frac{\partial (u_OX)}{\partial \alpha_a}]
U_O(u_OX\gets X) = U_O(u_OX\gets X)
\Omega_X [\frac{\partial X}{\partial \alpha_a}].
\l{5.17}
\eeb
The first  relation means that the action 1-form $\omega$ is conserved
under time evolution
\beq
\frac{d}{d\tau} \omega_{u_O^{\tau}X}
[\frac{\partial (u_O^{\tau}X)}{\partial \alpha_a}] = 0.
\l{5.18}
\eeq
Relation \r{5.18} is checked  by  a  direct  computation.  The  second
property means that the operator
$\Omega_{u_O^{\tau}X}
[\frac{\partial (u_O^{\tau}X)}{\partial \alpha_a}]$
takes solutions of
eq.\r{5.8} to solutions. This is true since
$\Omega_{u_O^{\tau}X}
[\frac{\partial (u_O^{\tau}X)}{\partial \alpha_a}]$
commutes with
$i \frac{d}{d\tau} -
\frac{1}{2} \Xi^2O(X^{\tau})   -   O^{(1)}(X^{\tau})$   according   to
eq.\r{5.10}.

\subsection{Semiclassical gauge transformations}

{\bf 1.} It has been noticed in section 3 that
quantum states
\beq
\Psi_H \sim \exp\left[
- \frac{i}{\sqrt{h}} \int d{\bf x} \alpha({\bf x}) \hat{\Lambda}_{\bf x}
\right] \Psi_H
\l{5.19}
\eeq
are gauge-equivalent   (for   the   Hamiltonian   gauge).   Therefore,
semiclassical states
\bez
K_{X_1}^h f_1 \sim
e^{-\frac{i}{\sqrt{h}}
\int d{\bf x} \alpha({\bf x}) \hat{\Lambda}_{\bf x}}
K_{X_1}^h f_1 \simeq
K^h_{u_{\Lambda[\alpha]}X}
U_{\Lambda[\alpha]}
(u_{\Lambda[\alpha]}X \gets X) f
\eez
are also gauge-equivalent.  Therefore,  for the Hamiltonian gauge, one
should introduce  an equivalence relation on the semiclassical bundle:
one should set
\bez
X_1\sim X_2 \Leftrightarrow X_2 = u_{\Lambda[\alpha]}X_1
\mbox{ for some $\alpha$};
\eez
moreover, $K_{X_1}^hf_1 \simeq K_{X_2}^hf_2$ iff
$f_2 = V(X_2\gets X_1)f_1$ for
\bez
V(X_2\gets X_1) = U_{\Lambda[\alpha]}(u_{\Lambda[\alpha]}X \gets X).
\eez

{\bf 2.}  For  the  Lorentz  gauge,  due the Gupta-Bleuler equivalence
relation, there is also a gauge transformation of another form
\bez
\Psi_L \sim
e^{-\frac{i}{\sqrt{h}} \int  d{\bf  x}  \kappa({\bf x}) \hat{E}_0({\bf
x})} \Psi_L.
\eez
For the Lorentz gauge, one should then set
\bez
X_1\sim X_2 \Leftrightarrow X_2 = u_{\Lambda[\alpha]}
u_{{\cal E}_0[\kappa]} X_1
\mbox{ for some $\alpha,\kappa$};
\eez
$K_{X_1}^hf_1 \simeq K_{X_2}^hf_2$ iff
$f_2 = V(X_2\gets X_1)f_1$ for
\bez
V(X_2\gets X_1) = U_{\Lambda[\alpha]}(u_{\Lambda[\alpha]}
u_{{\cal E}_0[\kappa]}X \gets u_{{\cal E}_0[\kappa]}X)
U_{{\cal E}_0[\kappa]}(u_{{\cal E}_0[\kappa]}X \gets X).
\eez

{\bf 3.} An explicit form of equivalence relation  is  the  following.
For the Hamiltonian gauge, property
$X_2 = u_{\Lambda[\alpha]}X_1$ means that
\beq
S^{(2)} = S^{(1)},
\quad
{\cal E}_k^{(2)} = {\cal E}_k^{(1)},
\quad
{\cal A}_k^{(2)} = {\cal A}_k^{(1)} - \partial_k \alpha,
\quad
\Pi^{(2)} = \Pi^{(1)} e^{i\alpha},
\quad
\Theta^{(2)} = \Theta^{(1)} e^{i\alpha}.
\l{5.24}
\eeq
The operator $V(X_2\gets X_1)$ is if the form
\beq
V(X_2\gets X_1) g[a^k,\vartheta,\vartheta^*] =
g[a^k,\vartheta e^{-i\alpha},\vartheta^*e^{i\alpha}].
\l{5.25}
\eeq

For the Lorentz gauge, equality
$X_2 =   u_{\Lambda[\alpha]}u_{{\cal   E}_0[\kappa]}X_1$  consists  of
relation ${\cal  A}_0^{(2)}  =  {\cal   A}_0^{(1)}   +   \kappa$   and
eqs.\r{5.24}, the operator $V(X_2\gets X_1)$ is
\beq
V(X_2\gets X_1) v[a^0,a^k,\vartheta,\vartheta^*] =
V(X_2\gets X_1) v[a^0,a^k,\vartheta e^{-i\alpha},\vartheta^*
e^{i\alpha}].
\l{5.26}
\eeq
It follows from relations \r{5.25}, \r{5.26} that
\beq
V(X_3\gets X_1) = V(X_3\gets X_2) V(X_2\gets X_1)
\l{5.27}
\eeq
for both gauges.

Notice also that properies \r{2.25} are partial cases of \r{5.17}.

\subsection{Semiclassical Poincare transformations}

To construct   semiclassical   Poincare   transformations   (classical
transformations $u_g:{\cal X} \to  {\cal  X}$  and  unitary  operators
$U_g(u_gX\gets X):  {\cal F}_X \to {\cal F}_{u_gX}$, notice that it is
possible to use decomposition \r{3|1} for the Poincare group. Thus, it
is sufficient   to  specify  semiclassical  spatial  translations  and
rotations, boosts and evolution.

The corresponding    1-parametric    subgroups     $g(\tau)     \equiv
(a_{\tau},\Lambda_{\tau})$ of  the  Poincare  group  are  presented in
table 3. For such cases,
\bez
u_{g(\tau)} \equiv u_O^{\tau},
\quad
U_{g(\tau)}(u_{g(\tau)}X \gets X) \equiv
U^{\tau}_O(u^{\tau}_O X \gets X).
\eez
Observables $O$  corresponding to 1-parametric subgroups are presented
in table 3. They indeed satidfy eq.\r{5.11}.

\begin{table}
\caption{
Equations for classical Poincare transformations
}
{\footnotesize
\begin{tabular}{|p{3cm}|p{13cm}|}
\hline
Element of Poincare group
$(a_{\tau},\Lambda_{\tau})$;
corresponding observable $\hat{O}$
&
Classical Poincare  transformation  $u_{a_{\tau},\Lambda_{\tau}}:  X^0
\mapsto X^{\tau}$ is found from classical equations
$\dot{\Phi} = \nabla_O\Phi$, $\dot{\Pi} = \nabla_O\Pi$,
$\dot{S} = \int d{\bf x} \Pi \dot{\Phi} - O(\Pi,\Phi)$
of  the form:
\\ \hline
$a_{\tau}=0$,
$\Lambda_{\tau} = \exp(\frac{\tau}{2}
l^{sm}\zeta_{sm})$;
$\zeta_{sm} = -\zeta_{ms}$,
spatial rotation;
$\hat{O} = - \frac{1}{2} {\cal M}^{lm} \zeta_{lm}$.
&
\begin{description}
\item[all gauges:]
\bey
\dot{\Theta}^{\tau} = \zeta_{kl} x^k\partial_l \Theta^{\tau};\quad
\dot{\Pi_{\theta}}^{\tau} = \zeta_{kl} x^k\partial_l \Pi_{\theta}^{\tau}; \quad
\dot{S}^{\tau} = 0; \\
\dot{\cal A}^{s \tau} = \zeta_{kl} x^k\partial_l {\cal A}^{s \tau}
+ \zeta_{sl} {\cal A}^{l\tau};\quad
\dot{\cal E}_s^{\tau} = \zeta_{kl} x^k\partial_l {\cal E}_s^{\tau}
+ \zeta_{sl} {\cal E}_l^{\tau}.
\eey
\item[Lorentz gauge:]
$\dot{\cal A}^{0 \tau} = \zeta_{kl} x^k\partial_l {\cal A}^{0 \tau}$.
\end{description}
\\ \hline
$a_{\tau}^0=0$, $a_{\tau}^k = b^k \tau$,
$\Lambda_{\tau} = 1$;
spatial translation;
$\hat{O} = b^k {\cal P}^k$.
&
\begin{description}
\item[all gauges:]
\bey
\dot{\Theta}^{\tau} = - b^k\partial_k \Theta^{\tau};\quad
\dot{\Pi_{\theta}}^{\tau} = - b^k\partial_k \Pi_{\theta}^{\tau};\quad
\dot{S}^{\tau} = 0; \\
\dot{\cal A}^{s\tau} = - b^k\partial_k {\cal A}^{s\tau};\quad
\dot{\cal E}_s^{\tau} = - b^k\partial_k {\cal E}_s^{\tau}.
\eey
\item[Lorentz gauge:]
$\dot{\cal A}^{0\tau} = - b^k\partial_k {\cal A}^{0\tau}$.
\end{description}
\\ \hline
$a_{\tau}^0 = -\tau$, $a^k_{\tau} = 0$,
$\Lambda_{\tau} = 1$;
evolution;
$\hat{O} = {\cal P}^0$.
&
\begin{description}
\item[all gauges:]
\bey
\dot{\Theta}^{\tau} = \Pi_{\theta}^{\tau} + i {\cal A}^{0\tau} \Theta^{\tau};\quad
-\dot{\Pi_{\theta}}^{\tau} =   -   D_iD_i   \Theta^{\tau}  +  m^2  \Theta^{\tau}
+
V'(\Theta^{\tau} \Theta^{\tau*})   \Theta^{\tau}   -    i{\cal    A}^{0\tau}
\Pi_{\theta}^{\tau};\\
\dot{\cal A}^{k\tau}  =   {\cal   E}_k^{\tau}   -   \partial_k   {\cal
A}^{0\tau};\quad
-\dot{\cal E}_k^{\tau} = i[(D_k\Theta^{\tau})^* \Theta^{\tau} - \Theta^{\tau
*} D_k \Theta^{\tau}] - \partial_j (\partial_j{\cal  A}^k  -  \partial_k
{\cal A}^j);\\
\dot{S}^{\tau} = \int d{\bf x} [...];\\
D_k \equiv \partial_k + i {\cal A}^{k\tau}.
\eey
\item[Hamiltonian gauge:] ${\cal A}^{0\tau} = 0$;
\item[Couloumb gauge:]       ${\cal       A}^{0\tau}       \Rightarrow
\frac{1}{\partial^2} \partial_l {\cal E}_l^{\tau}$;
\item[Lorentz gauge:]  $\dot{\cal  A}^{0\tau}  =  -  \partial_k  {\cal
A}^{k\tau}$.
\end{description}
\\ \hline
$a_{\tau}=0$,
$\Lambda_{\tau} = \exp(-{\tau}
n^k l^{k0})$;
boost;
$\hat{O} = n^k {\cal M}^{k0}$.
&
\begin{description}
\item[all gauges:]
\bey
\dot{\Theta}^{\tau} =   n^sx^s   [\Pi_{\theta}^{\tau}   +   i   {\cal  A}^{0\tau}
\Theta^{\tau}];\quad
\dot{\cal A}^{k\tau}  =  x^sn^s {\cal   E}_k^{\tau}
-   \partial_k (x^sn^s  {\cal A}^{0\tau});\\
-\dot{\Pi_{\theta}}^{\tau} =   -   D_i x^sn^sD_i   \Theta^{\tau}
+  x^sn^s (m^2 \Theta^{\tau}
+
V'(\Theta^{\tau} \Theta^{\tau*})  \Theta^{\tau}
-    i  {\cal    A}^{0\tau} \Pi_{\theta}^{\tau});\\
-\dot{\cal E}_k^{\tau} = ix^sn^s
[(D_k\Theta^{\tau})^* \Theta^{\tau} - \Theta^{\tau
*} D_k \Theta^{\tau}] - \partial_j x^sn^s
(\partial_j{\cal  A}^k  -  \partial_k {\cal A}^j);\\
\dot{S}^{\tau} = \int d{\bf x} [...];\\
D_k \equiv \partial_k + i {\cal A}^{k\tau}.
\eey
\item[Hamiltonian gauge:]
${\cal A}^{0\tau} = 0$;
\item[Couloumb gauge:]
$x^sn^s {\cal  A}^{0\tau}
\Rightarrow  \frac{1}{\partial^2} \partial_l x^sn^s {\cal E}_l^{\tau}$;
\item[Lorentz gauge:]
$\dot{\cal A}^{0\tau} = - \partial_k x^sn^s {\cal A}^{k\tau}$.
\end{description}
\\ \hline
\end{tabular}
}
\end{table}

Notice that   properties   \r{2.17}    of    semiclassical    Poincare
transformations are partial cases of \r{5.17}.

\subsection{Manifestly covariant semiclassical observables and fields}

{\bf 1.}   In   the  previous  subsections,  we  have  considered  the
semiclassical field operators in the Hamiltonian framework. The fields
depended on the spatial coordinates only.

Let us  consider  now the Poincare covariant observables.  They should
depend on Heisenberg fields
$\hat{\varphi}(x)
= (\hat{A}^{\mu}(x),\hat{\theta}(x),\hat{\theta}^*(x))$:
\beq
\hat{O} = O(\sqrt{h}\hat{\varphi}(\cdot)) =
O(\sqrt{h}\hat{A}^{\mu}(\cdot),\sqrt{h}\hat{\theta}(\cdot),
\sqrt{h}\hat{\theta}^*(\cdot)).
\l{5a.1}
\eeq
For gauge   theories,   only  gauge-invariant  observables  should  be
considered. This means that quantum expression \r{5a.1}  specidies  an
observable iff       the       classical      functional
$O({\cal A}^{\mu}(\cdot),\Theta(\cdot),\Theta^*(\cdot))$
is   invariant   under
gauge transformations
\beq
O({\cal A}^{\mu} + \partial_{\mu} \alpha,
\Theta e^{i\alpha},\Theta^* e^{-i\alpha}) =
O({\cal A}^{\mu}, \Theta,\Theta^*).
\l{5a.2}
\eeq
One can  rewrite property \r{5a.2} in the infinitesimal form.  Namely,
expanding the left-hand side of relation \r{5a.2} in  $\alpha(\cdot)$,
one finds that
\bez
\int dx
\left[
\frac{\delta O}{\delta {\cal A}^{\mu}(x)} \partial_{\mu} \alpha(x)
+ \frac{\delta O}{\delta \Theta(x)} i\alpha(x) \Theta(x)
+ \frac{\delta O}{\delta \Theta^*(x)} (- i\alpha(x) \Theta(x))
\right] = 0.
\eez
This means that
\beq
\partial_{\mu}
\frac{\delta O}{\delta {\cal A}^{\mu}(x)} =
i\left[
\Theta(x) \frac{\delta O}{\delta \Theta(x)}
- \Theta^*(x) \frac{\delta O}{\delta \Theta^*(x)}
\right].
\l{5a.3}
\eeq
A formal analog of commutation rule \r{2.7} for the field is
\bez
\hat{O} K^h_X f \simeq K^h_X
O({\cal A}^{\mu} + \sqrt{h} \hat{a}^{\mu},
\Theta + \sqrt{h}\hat{\vartheta}, \Theta^* + \sqrt{h}\hat{\vartheta}^*)
f \simeq
K^h_X [O(X) + \sqrt{h} \Xi O(X)]f
\eez
with
\beb
O(X) \equiv O({\cal A}^{\mu}(\cdot),\Theta(\cdot),\Theta^*(\cdot)),\\
\Xi O(X) \equiv
\int dx
\left(
\frac{\delta O}{\delta {\cal A}^{\mu}(x)} \hat{a}^{\mu}(x|X) +
\frac{\delta O}{\delta \Theta (x)} \hat{\vartheta}(x|X) +
\frac{\delta O}{\delta \Theta^* (x)} \hat{\vartheta}^*(x|X)
\right).
\l{5a.4}
\eeb
Therefore, there is  the  following  specific  feature  of  the  gauge
theory. For    the    scalar   field   theory,   semiclassical   field
$\hat{\phi}(x|X)$ is  a  well-defined  operator  distribution  in  the
following sense:  expression  $\int  dx  \hat{\phi}(x|X)  \frac{\delta
O}{\delta \Phi(x)}$ specifies a well-defined operator for  any  smooth
rapidly damping   at   infinity   function   $\frac{\delta   O}{\delta
\Phi(x)}$. For the electrodynamic case, the linear combination \r{5.4}
should specify  a  well-defined  operator,  provided that the c-number
functions
$\frac{\delta O}{\delta {\cal A}^{\mu}(x)}$,
$\frac{\delta O}{\delta \Theta (x)}$,
$\frac{\delta O}{\delta \Theta^* (x)}$ satisfy eq.\r{5a.3}.

To define  c-number  quantity  $O(X)$ and operator $\Xi O(X)$,  let us
introduce manifestly covariant notations.

{\bf 2.} Let us identify elements $X\in \cal X$ with sets
$\overline{X} = (S,\overline{\Phi}(x)) \equiv
(S,\overline{\cal A}^{\mu}(x),\overline{\Theta}(x),
\overline{\Theta}^*(x)) \in \overline{\cal X} = \{(
\overline{X} )\}$ analogously to section 2. Here
$\overline{\Phi}(x) \equiv  \overline{\Phi}(x|X)$  is  a  solution  of
system of classical equations
\beb
\partial_{\nu} \overline{F}^{\mu\nu} =
i (\overline{\Theta}^* \overline{D}^{\mu} \overline{\Theta}
- \overline{\Theta} \overline{D}^{\mu} \overline{\Theta}^*);
\\
\overline{D}_{\mu} \overline{D}^{\mu} \overline{\Theta}
+ m^2 \overline{\Theta} + V'(\overline{\Theta}^* \overline{\Theta})
\overline{\Theta} = 0
\l{5a.5}
\eeb
with
\bez
\overline{\cal F}^{\mu \nu} =
\partial^{\mu} \overline{\cal A}^{\nu}
- \partial^{\nu} \overline{\cal A}^{\mu},
\qquad
\overline{D}_{\mu} = \partial_{\mu} - i\overline{\cal A}_{\mu}.
\eez
Initial conditions for system \r{5a.5} are as follows:
\beb
\overline{\Theta}|_{x^0=0} = \Theta({\bf x}),
\quad
\overline{D}_0 \overline{\Theta}|_{x^0=0} = \Pi_{\theta}({\bf x}),
\\
\overline{\cal A}^k|_{x^0=0} = {\cal A}^k({\bf x}),
\quad
\overline{F}^{0k}|_{x^0=0} = {\cal E}_k({\bf x}),
\quad
\overline{\cal A}^0|_{x^0=0} = {\cal A}^0({\bf x}).
\l{5a.6}
\eeb
Condition for
$\overline{\cal A}^0$ should be imposed for Lorentz gauge only.

It is well-known that a  solution  to  the  Cauchy  problem  \r{5a.5},
\r{5a.6} is  defined  up  to  a  gauge transformation.  Namely,  if we
constructed one of solutions
$(\overline{\cal A}^{\mu}(x),\overline{\Theta}(x))$ then the functions
\beq
\overline{\cal A}^{\mu}(x) + \partial_{\mu}\rho(x),
\qquad
\overline{\Theta}(x) e^{i\rho(x)}
\l{5a.7}
\eeq
would also satisfy system \r{5a.5}.  For different  gauges,  different
additional gauge conditions are to be imposed then.

Making use of the introduced notations, set
\bez
O(X) \equiv O[\overline{\Phi}(\cdot)].
\eez
This is   a  well-defined  expression  since  the  functional  $O$  is
invariant under gauge transformations \r{5a.7}.

For $X_1\sim X_2$, one also checks property \r{2.35}
\bez
O(X_2) = O(X_1),
\eez
since gauge-equivalent initial conditions for system \r{5a.5} generate
gauge-equivalent solutions.

Let us check now property \r{2.30}. It can be written as
\beq
O(\overline{\Phi}(\cdot|u_gX)) =
(v_gO)(\overline{\Phi}(\cdot|X)) \equiv
O(v_g\overline{\Phi}(\cdot|X)).
\l{5a.8}
\eeq
Property \r{5a.8} means that the space-time functions
\beq
\tilde{\cal A}^{\mu}(x)       =       \Lambda^{\mu}_{\nu}        {\cal
A}^{\nu}(\Lambda^{-1}(x-a));
\qquad
\tilde{\Theta}(x) = \Theta (\Lambda^{-1}(x-a))
\l{5a.9}
\eeq
satisfies system \r{5a.5}, while initial conditions
$\tilde{X} =             (\tilde{\cal             E}_{\mu},\tilde{\cal
A}^{\mu},\tilde{\Pi}_{\theta},\tilde{\Theta})$
\beb
\tilde{\Theta}({\bf x}) \equiv \tilde{\Theta}|_{x^0=0},
\quad
\tilde{\Pi_{\theta}}({\bf x}) \equiv
\tilde{D}_0 \tilde{\Theta}|_{x^0=0},
\\
\tilde{\cal A}^k({\bf x}) \equiv \tilde{\cal A}^k|_{x^0=0},
\quad
\tilde{\cal E}_k({\bf x}) \equiv \tilde{\cal F}^{0k}|_{x^0=0}
\l{5a.10}
\eeb
are gauge-equivalent to $u_gX$:
\beq
\tilde{X} \sim u_gX.
\l{5a.11}
\eeq
System \r{5a.5}  for  functions  \r{5a.9} is satisfied due to Poincare
invariance of  \r{5a.5}l  property  \r{5a.11}  is  checked  by  direct
calculations for  partial  cases:  spatial translations and rotations,
evolution and boosts.

{\bf 3.} A tangent vector $\delta X \in T{\cal X}$ can  be  identified
with a set $\delta \overline{X} \equiv
(\delta \overline{S},\delta \overline{\Phi}(x))
\in T\overline{\cal X}$;  $\delta \overline{\Phi}$ being a solution to
the variation system. Analogously to \r{2.18},
\beb
\delta \{
\partial_{\nu} \overline{F}^{\mu\nu} -
i (\overline{\Theta}^* \overline{D}^{\mu} \overline{\Theta}
- \overline{\Theta} \overline{D}^{\mu} \overline{\Theta}^*)
\} = 0;
\\
\delta\{
\overline{D}_{\mu} \overline{D}^{\mu} \overline{\Theta}
+ m^2 \overline{\Theta} + V'(\overline{\Theta}^* \overline{\Theta})
\overline{\Theta}\} = 0.
\l{5a.12}
\eeb
Then one  introduces  the  operator  $\Omega[\delta\overline{\Phi}]  =
\Omega[\delta X]$.  Notice that correspondence $(\delta\Pi,\delta\Phi)
\mapsto \delta \overline{\Phi}$ is not one-to-one;  however,  one  has
$\Omega[\delta X] = 0$ if $X+\delta X \sim X$; therefore, the operator
$\Omega[\delta\overline{\Phi}]$ is well-defined.

The commutation     relation      \r{2.14}      between      operators
$\Omega[\delta\overline{\Phi}]$ can be taken to a manifestly covariant
form. Making use of eq.\r{4.6g}, one obtains
\bey
[\Omega[\delta_1\overline{\Phi}],\Omega[\delta_2\overline{\Phi}]] =
\\
- i \int_{x^0=0} d{\bf x}
[\delta_1{\cal E}_k  \delta_2  {\cal  A}^k  +
\delta_1   \Pi_{\theta} \delta_2 \Phi^* +
\delta_1   \Pi_{\theta}^* \delta_2 \Phi -
\delta_2{\cal E}_k  \delta_1  {\cal  A}^k  -
\delta_2   \Pi_{\theta} \delta_1 \Phi^* -
\delta_2   \Pi_{\theta}^* \delta_1 \Phi] =
\\
-i \int_{x^0=0} d{\bf x}
\left[
\delta_1 \frac{\partial \cal L}{\partial \overline{\Phi}_{,0}}
\delta_2 \overline{\Phi} -
\delta_2 \frac{\partial \cal L}{\partial \overline{\Phi}_{,0}}
\delta_1 \overline{\Phi}
\right]
\eey
with
$\overline{\Phi} \equiv
(\overline{\cal A}^{\mu},\overline{\Theta},\overline{\Theta}^*)$,
the notation     $\overline{\Phi}_{,\mu}     \equiv     \partial_{\mu}
\overline{\Phi}$ is introduced,
$\cal L$ is the classical Lagrangian
\bez
{\cal L} =
\overline{D}_{\mu} \overline{\Theta}^*
\overline{D}^{\mu} \overline{\Theta} -
m^2 \overline{\Theta}^* \overline{\Theta} -
V(\overline{\Theta}^* \overline{\Theta}) -
\frac{1}{4} \overline{\cal F}_{\mu\nu} \overline{\cal F}^{\mu\nu}.
\eez
One can notice that
\bez
\partial_{\mu}
\left[
\delta_1 \frac{\partial \cal L}{\partial \overline{\Phi}_{,\mu}}
\delta_2 \overline{\Phi} -
\delta_2 \frac{\partial \cal L}{\partial \overline{\Phi}_{,\mu}}
\delta_1 \overline{\Phi}
\right] = 0.
\eez
Therefore, the  commutator  relation  is  taken  to   the   manifestly
covariant form
\beq
[\Omega[\delta_1\overline{\Phi}],\Omega[\delta_2\overline{\Phi}]] =
- i\int d\sigma_{\mu}
\left[
\delta_1 \frac{\partial \cal L}{\partial \overline{\Phi}_{,\mu}}
\delta_2 \overline{\Phi} -
\delta_2 \frac{\partial \cal L}{\partial \overline{\Phi}_{,\mu}}
\delta_1 \overline{\Phi}
\right]
\l{5a.13}
\eeq

{\bf 4.} Let us express the  operator  $\Xi  O(X)$  via  the  operator
1-form $\Omega$.   Since   $\Xi  O(X)$  is  a  linear  combination  of
semiclassical fields, it should be of the form
\beq
\Xi O(X) = - \Omega [\nabla_O\overline{\Phi}].
\l{5a.14}
\eeq
Let us find an explicit form of $\nabla_O\overline{\Phi}$.  It  should
be obtained from relation \r{2.34}:
\beq
\delta O = \int d\sigma_{\mu}
\left[
\delta \frac{\partial \cal L}{\partial \overline{\Phi}_{,\mu}}
\nabla_O \overline{\Phi} -
\nabla_O \frac{\partial \cal L}{\partial \overline{\Phi}_{,\mu}}
\delta \overline{\Phi}
\right].
\l{5a.15}
\eeq
One can construct variation $\nabla_O\overline{\Phi}$ in the following
way. First, consider the function
$\overline{\nabla_O}\overline{\Phi}$ satisfying classical equations of
motion with an external source:
\beb
\overline{\nabla_O}
\left\{
\frac{\partial \cal L}{\partial \overline{\Phi}}
-
\left(\frac{\partial \cal L}{\partial \overline{\Phi}_{,\alpha}}
\right)_{,\alpha}
\right\} =
\frac{\delta O}{\delta \overline{\Phi}}, \\
\overline{\nabla_O} \overline{\Phi}|_{x^0\to - \infty} = 0.
\l{5a.16}
\eeb
Then, let $\nabla_O\Phi$ be a solution of variation system
with boundary condition at $+\infty$:
\beb
{\nabla_O}
\left\{
\frac{\partial \cal L}{\partial \overline{\Phi}}
-
\left(\frac{\partial \cal L}{\partial \overline{\Phi}_{,\alpha}}
\right)_{,\alpha}
\right\} = 0,
\\
\nabla_O\overline{\Phi} \equiv
\nabla_O \overline{\Phi}|_{x^0=+\infty}.
\l{5a.16a}
\eeb
If the variation $\frac{\delta  O}{\delta  \overline{\Phi}(x)}$  is  a
function with  compact  support  (the  observable  $O$ is local),  the
limits $x^0\to \pm\infty$ mean $x> supp
\frac{\delta  O}{\delta  \overline{\Phi}(x)}$ and $x<
\frac{\delta  O}{\delta  \overline{\Phi}(x)}$.

Let us check eq.\r{5a.15}. One takes the right-hand side to the form:
\beb
\int_{x^0 \to + \infty} d\sigma_{\mu}
\left[
\delta \frac{\partial \cal L}{\partial \overline{\Phi}_{,\mu}}
\overline{\nabla_O} \overline{\Phi} -
\overline{\nabla_O} \frac{\partial \cal L}{\partial \overline{\Phi}_{,\mu}}
\delta \overline{\Phi}
\right] =
\int dx
\partial_{\mu}
\left[
\delta \frac{\partial \cal L}{\partial \overline{\Phi}_{,\mu}}
\overline{\nabla_O} \overline{\Phi} -
\overline{\nabla_O} \frac{\partial \cal L}{\partial \overline{\Phi}_{,\mu}}
\delta \overline{\Phi}
\right] =
\\
\int dx
\left[
\delta
\left(
\frac{\partial \cal L}{\partial \overline{\Phi}_{,\mu}}
\right)_{,\mu}
\overline{\nabla_O} \overline{\Phi} +
\delta
\frac{\partial \cal L}{\partial \overline{\Phi}_{,\mu}}
(\overline{\nabla_O} \overline{\Phi})_{,\mu}  -
\overline{\nabla_O}
\left(
\frac{\partial \cal L}{\partial \overline{\Phi}_{,\mu}}
\right)_{,\mu}
\delta \overline{\Phi}
-
\overline{\nabla_O} \frac{\partial \cal L}{\partial \overline{\Phi}_{,\mu}}
(\delta \overline{\Phi})_{,\mu}
\right].
\l{5a.17}
\eeb
Making use of property \r{5a.16} and relation
\bez
\delta
\left\{
\frac{\partial \cal L}{\partial \overline{\Phi}}
-
\left(\frac{\partial \cal L}{\partial \overline{\Phi}_{,\alpha}}
\right)_{,\alpha}
\right\} = 0,
\eez
one takes the right-hand side of expression \r{5a.17} to the form
\bez
\int dx     \frac{\delta    O}{\delta    \overline{\Phi}(x)}    \delta
\overline{\Phi}(x).
\eez
This coincides with the left-hand side $\delta O$.  Therefore, formula
\r{5a.15} is  satisfied  and  $\Xi  O(X)$ has the form \r{5a.14} under
conditions \r{5a.16}  and  \r{5a.16a}.  This  is  in  agreement   with
analogous relations for scalar field theory (section 2).

An explicit form of eqs.\r{5a.16} for scalar electrodynamics is
\beb
\overline{\nabla_O}\{
\partial_{\nu} \overline{F}^{\mu\nu} -
i (\overline{\Theta}^* \overline{D}^{\mu} \overline{\Theta}
- \overline{\Theta} \overline{D}^{\mu} \overline{\Theta}^*)
\} =  - \frac{\delta O}{\delta \overline{\cal A}_{\mu}(x)};
\\
\overline{\nabla_O}\{
\overline{D}_{\mu} \overline{D}^{\mu} \overline{\Theta}
+ m^2 \overline{\Theta} + V'(\overline{\Theta}^* \overline{\Theta})
\overline{\Theta}\} =
\frac{\delta O}{\delta \overline{\Theta}^*(x)};
\\
\overline{\nabla_O} \{
\overline{D}_{\mu} \overline{D}^{\mu} \overline{\Theta}^*
+ m^2 \overline{\Theta}^* + V'(\overline{\Theta}^* \overline{\Theta})
\overline{\Theta}^* \} =
\frac{\delta O}{\delta \overline{\Theta}(x)};
\\
\overline{\nabla_O} \overline{\cal A}^{\mu}|_{x^0\to-\infty} = 0,
\quad
\overline{\nabla_O} \overline{\Theta}|_{x^0\to-\infty} = 0,
\quad
\overline{\nabla_O} \overline{\Theta}^*|_{x^0\to-\infty} = 0
\l{5a.18}
\eeb
and
\bey
{\nabla_O} \overline{\cal A}^{\mu}|_{x^0\to +\infty} =
\overline{\nabla_O} \overline{\cal A}^{\mu}|_{x^0\to +\infty},
\\
{\nabla_O} \overline{\Theta}|_{x^0\to +\infty} =
\overline{\nabla_O} \overline{\Theta}|_{x^0\to +\infty},
\quad
{\nabla_O} \overline{\Theta}^*|_{x^0\to +\infty} =
\overline{\nabla_O} \overline{\Theta}^*|_{x^0\to +\infty}.
\eey
Notice that  variation  system  \r{5a.18}  is  solvable  iff the gauge
invariance condition \r{5a.3} is satisfied.

{\bf 5.} One can now check that relations \r{2.30}  and  \r{2.35}  for
$\Xi O$  are  indeed  satisfied.  Property  \r{2.30} is a corollary of
Poincare invariance of system \r{5a.18} and analogous property for the
operators $\Omega$.   Property   \r{2.35}  is  a  corollary  of  gauge
invariance of  the  observable  $O$  and  property  \r{2.25}  for  the
operators $\Omega$.

\subsection{Conservation of gauge equivalence relation}

An important  property  of  semiclassical  Poincare transformations is
that they should take gauge equivalent semiclassical states  to  gauge
equivalent (eq.\r{2.24}).  Let  us  investigate  this property for the
more general case of semiclassical transformation:
\beq
(X^0,f^0) \mapsto (X^{\tau} = u_O^{\tau}X^0,
f^{\tau} = U^{\tau}_O (u^{\tau}_O X \gets X) f^0).
\l{5c.1a}
\eeq
One should check whether
\bez
(X_1,f_1) \sim (X_2,f_2)
\eez
implies that
\beq
(u^{\tau}_O X_1, U^{\tau}_O (u^{\tau}_OX_1 \gets X_1) f_1)
\sim
(u^{\tau}_O X_2, U^{\tau}_O (u^{\tau}_OX_2 \gets X_2) f_2).
\l{5c.1}
\eeq
Property \r{5c.1} is not convenient for check.  It is more suitable to
consider gauge-invariant sections  of  the  semiclassical  bundle.  To
specify a  section  $\chi$,  one should assign a quantum state $\chi_X
\in {\cal F}_X$ to each $X\in {\cal X}$. We say that section $\chi$ is
gauge invariant iff
\beq
\chi_{X_2} = V(X_2\gets X_1) \chi_{X_1}
\l{5c.2}
\eeq
for all $X_1 \sim X_2$.

It is remarkable that semiclassical states $(X_1,f_1)$ and $(X_2,f_2)$
are gauge  equivalent  iff for all gauge invariant sections $\chi$ the
relation
\beq
(\chi_{X_1},f_1) = (\chi_{X_2},f_2)
\l{5c.3}
\eeq
is satisfied.  Eq.\r{5c.3}  is  a  convenient necessary and sufficient
condition of equivalence of semiclassical states.

An automorphism \r{5c.1a} of semiclassical bundle  may  be  viewed  as
transformation in the space of sections. Namely, consider the operator
$\check{U}^{\tau}_O$ taking section $\chi$ to the following section
\beq
(\check{U}^{\tau}_O\chi)_X =
U^{\tau}_O (X\gets u^{-\tau}_OX) \chi_{u^{-\tau}_OX}.
\l{5c.4}
\eeq
It happens that property \r{5c.1} means that gauge invariant  sections
are taken  to  gauge invariant.  This can be checked by using identity
\r{5c.3}.

To justify that the property of gauge invariance of section $\chi$  is
conserved under  time  evolution,  one  can  notice  that  the section
$\chi_O^{\tau} \equiv \check{U}_O^{\tau} \chi$ satisfies the equation
\bez
i\frac{d}{d\tau} \chi^{\tau}_O = \check{O} \chi^{\tau}_O
\eez
with
\bez
\check{O} = \frac{1}{2} \Xi^2 O(X) + O^{(1)}(X) - i\nabla_O.
\eez

Condition of gauge invariance of the section can be written as
\beb
\check{\Lambda}[\alpha] \chi^{\tau}_O = 0,
\quad \mbox{ Hamiltonian gauge};\\
\check{\Lambda}[\alpha] \chi^{\tau}_O = 0, \mbox{ and }
\check{\cal E}_0[\kappa] \chi^{\tau}_O = 0,
\quad \mbox{ Lorentz gauge. }
\l{5c.5}
\eeb
Therefore, one should check that
\beq
[\check{O}, \check{\Lambda}[\alpha]] \chi^{\tau}_O = 0,
\quad
[\check{O}, \check{\cal E}_0[\kappa]] \chi^{\tau}_O = 0
\l{5c.6}
\eeq
under conditions \r{5c.5}.

It is shown in ref. \c{Shvedov4} that for classical observables $A$ and $B$
\beb
[
i\nabla_A - \frac{1}{2} \Xi^2 A - A^{(1)},
i\nabla_B - \frac{1}{2} \Xi^2 B - B^{(1)}] =
\\
i
(i\nabla_{\{A;B\}} - \frac{1}{2} \Xi^2 \{A;B\}
+ \nabla_B A^{(1)} - \nabla_A B^{(1)}),
\l{5c.6a}
\eeb
provided that  the  Weyl  quantization  is  used.  Here $\{A;B\}$ is a
Poisson bracket. Thus, for the case of gauge-invariant observables,
\beq
\{ O; \Lambda[\alpha]\} = 0,
\qquad
\{ O; {\cal E}_0 [\kappa] \} = 0
\l{5c.7}
\eeq
on the constraint surface, properties \r{5c.6} are formally satisfied.
However, one should be careful with quantum corrections $O^{(1)}$  due
to divergences and renormalization.

For the Poincare generators, one has:
\begin{itemize}
\item for the Hamiltonian gauge,
\bey
\{ \Lambda_{\bf x} ; {\cal H}\} = 0, \qquad
\{ \Lambda_{\bf x} ; {\cal M}^{k0}\} = 0, \\
\{ \Lambda_{\bf x} ; {\cal P}^l \} = \partial_l \Lambda_{\bf x},
\qquad
\{ \Lambda_{\bf x} ; {\cal M}^{kl}\} =
(x^k\partial_l - x^l\partial_k) \Lambda_{\bf x};
\eey
\item for the Lorentz gauge,
$\{ \Lambda_{\bf x} ; {\cal P}^l \}$ and
$\{ \Lambda_{\bf x} ; {\cal M}^{kl}\}$ are the same,
while
\bey
\{ \Lambda_{\bf x}; {\cal H} \} = \Delta_{\bf x} {\cal E}_0({\bf x});
\qquad
\{ \lambda_{\bf x}; {\cal M}^{k0}\} =
\partial_s x^k \partial_s {\cal E}_0({\bf x});
\\
\{ {\cal E}_0({\bf x}); {\cal H} \} = \Lambda_{\bf x};
\qquad
\{ {\cal E}_0({\bf x}); {\cal M}^{k0}\} = x^k \Lambda_{\bf x};
\\
\{ {\cal E}_0({\bf x}); {\cal P}^l\} =
\partial_l {\cal E}_0({\bf x});
\qquad
\{ {\cal E}_0({\bf x}); {\cal M}^{kl} \}
= (x^k \partial_l - x^l \partial_k) {\cal E}_0({\bf x}).
\eey
\end{itemize}
Relations \r{5c.7} are satisfied.

\subsection{On group and infinitesimal properties}

The remaining  property  to  be  checked  is  eq.\r{2.26}.  It  can be
simplified. Consider the operator $\check{U}_g$ in the space of  gauge
invariant sectons $\chi$:
\beq
(\check{U}_g \chi)_X = U_g(X\gets u_{g^{-1}}X) \chi_{u_{g^{-1}}X}.
\l{5c.8}
\eeq
Relation \r{2.26} will be rewritten as
\beq
\check{U}_{g_1g_2} \chi = \check{U}_{g_1} \check{U}_{g_2} \chi.
\l{5c.9}
\eeq
Therefore, the  correspondence $g \mapsto \check{U}_g$ in the space of
sections is  a  representation  of  the  Poincare  group.   To   check
eq.\r{5c.9}, it is more convenient to justify infinitesimal analogs of
\r{5c.9}:
\beb
\left[\check{P}^{\lambda}; \check{P}^{\mu}\right] \chi = 0;
\qquad
\left[\check{M}^{\lambda\mu}; \check{P}^{\sigma}\right] \chi
= i    (g^{\mu\sigma}    \check{P}^{\lambda}    -    g^{\lambda\sigma}
\check{P}^{\mu}) \chi; \\
\left[\check{M}^{\lambda\mu}; \check{M}^{\rho\sigma}\right] =
- i
(g^{\lambda \rho} \check{M}^{\mu \sigma}
- g^{\lambda \sigma} \check{M}^{\mu \rho}
+ g^{\mu \sigma} \check{M}^{\lambda \rho}
- g^{\mu \rho} \check{M}^{\lambda \sigma}
) \chi
\l{5c.10}
\eeb
under conditions \r{5c.5}.  making use  of  relations  \r{5c.6a},  one
reduces relation \r{5c.10} to the classical formulas:
\beb
\{ {\cal P}^{\lambda}; {\cal P}^{\mu}\} = 0;
\qquad
\{ {\cal M}^{\lambda\mu}; {\cal P}^{\sigma}\}
= i    (g^{\mu\sigma}    {\cal P}^{\lambda}    -    g^{\lambda\sigma}
{\cal P}^{\mu}); \\
\{{\cal M}^{\lambda\mu}; {\cal M}^{\rho\sigma}\} =
- i
(g^{\lambda \rho} {\cal M}^{\mu \sigma}
- g^{\lambda \sigma} {\cal M}^{\mu \rho}
+ g^{\mu \sigma} {\cal M}^{\lambda \rho}
- g^{\mu \rho} {\cal M}^{\lambda \sigma}
).
\l{5c.11}
\eeb
For the Lorentz gauge,  relations \r{5c.11} are satisfied exactly, for
the Hamiltonian gauge,  they are valid on constraint surface.  For the
Couloumb gauge, one can reduce it to one of other gauges.

\section{Conclusions}

Thus, axioms   of   semiclassical  scalar  electrodynamics  have  been
discussed. The considered approach is  not  manifestly  covariant,  so
that a rigorous proof of properties of semiclassical theory (analog of
\c{Shvedov1}) is not easy.

It is possible to simplify the semiclassical theory.  One should use a
manifestly covariant  semiclassical  approach  \c{Shvedov6}.  For this
approach, axioms formulated  here  are  also  valid;  however,  it  is
BRST-BFV quantization that can be formulated in this way; on the other
hand, the  Hamiltonian  approach  of  this  paper  is  applicable   to
Hamiltonian and Couloumb gauges as well.

One can   also   investigate   the  semiclassical  properties  of  the
non-abelian gauge theories.

The author is going to clarify these problems in further publications.

\end{document}